\documentclass[preprint,12pt]{elsarticle}

\usepackage{amssymb}
\usepackage{amsmath}

\usepackage{array}
\usepackage[caption=false,font=normalsize,labelfont=sf,textfont=sf]{subfig}
\usepackage{textcomp}
\usepackage{stfloats}
\usepackage{url}
\usepackage{verbatim}
\usepackage{graphicx}
\usepackage{multirow}
\usepackage{color}
\usepackage{wasysym}
\usepackage{pifont}
\usepackage{diagbox}
\usepackage{threeparttable}
\usepackage{makecell}
\usepackage{tabularx}
\usepackage{longtable}

\usepackage{stackengine}
\newcommand\xrowht[2][0]{\addstackgap[.5\dimexpr#2\relax]{\vphantom{#1}}}

\journal{}

\begin{document}

\begin{frontmatter}



\title{Edge Computing for IoT: Novel Insights from a Comparative Analysis of Access Control Models}


\author[a]{Tao Xue}
\author[a]{Ying Zhang}
\author[a]{Yanbin Wang}
\author[b]{Wenbo Wang}
\author[b]{Shuailou Li}
\author[c]{Haibin Zhang} 

\affiliation[a]{organization={The Hangzhou Institute of Technology,  Xidian University},
            city={Hangzhou},
            postcode={311231}, 
            country={China}}
\affiliation[b]{organization={The School of Cyber Security, University of Chinese Academy of Sciences},
            city={Beijing},
            postcode={100085}, 
            country={China}}
\affiliation[c]{organization = {The School of Cyber Security, Xidian University},
addressline={No.8 East Qiannong Road, Xiaoshan District, Hangzhou City, Zhejiang Province}, 
            postcode={311231}, 
            country={China, Corresponding author}}
\begin{abstract}

IoT edge computing positions computing resources closer to the data sources to reduce the latency, relieve the bandwidth pressure on the cloud, and enhance data security. Nevertheless, data security in IoT edge computing still faces critical threats (e.g., data breaches). Access control is fundamental for mitigating these threats. However, IoT edge computing introduces notable challenges for achieving resource-conserving, low-latency, flexible, and scalable access control. To review recent access control measures, we novelly organize them according to different data lifecycles—data collection, storage, and usage—and, meanwhile, review blockchain technology in this novel organization. In this way, we provide novel insights and envisage several potential research directions. This survey can help readers find gaps systematically and prompt the development of access control techniques in IoT edge computing under the intricacy of innovations in access control.
\end{abstract}


\end{frontmatter}



\section{Introduction}
\label{Sec::Introduction}

In the Internet of Things (IoT) era, massive devices generate enormous data, which is collected, stored, and analyzed for fueling many contemporary applications, such as smart homes and smart cities \cite{gomez2019internet}. 
As the pivotal infrastructure, cloud computing is used to harness these data. 
However, the latency, mainly induced by the geographical distance between data sources and cloud centers, would be unacceptable for some applications, such as virtual realization and connected autonomous vehicles.   
To address these challenges, edge computing acts as a synergistic technique to cloud computing, so that computing resources are situated closer to data sources and large raw IoT data is processed locally in edge nodes. Subsequently, the processed data could be transmitted to the cloud for advanced analysis \cite{chen2019deep}. 
Therefore, edge computing can help alleviate network burdens for the cloud, support latency-sensitive applications, and simultaneously enhance data security \cite{shi2016edge}. Nevertheless, data security in IoT edge computing faces various threats, including unauthorized access, data breaches, etc. which can lead to privacy violations and service disruptions.

The threats to data security in IoT edge computing stem from various sources \cite{alwarafy2020survey, xiao2019edge}. For instance, the large attack surface due to the multitude of IoT devices, many of which lack robust security controls, makes the system vulnerable to attacks. Additionally, the distributed nature of edge computing can lead to security inconsistencies across different nodes, and the proximity to data sources can make sensitive information more accessible to potential attackers.
Access control plays a critical role in mitigating these threats. By implementing robust access control measures, it is possible to ensure that only authorized entities have access to the data, thereby reducing the risk of unauthorized access and data breaches. Access control mechanisms (e.g., authorization, and encryption) can help in safeguarding the data at rest and in transit.

These mechanisms depend on many typical and popular access control models, such as Capability-Based Access Control (CapBAC) \cite{hernandez2013distributed}, Role-Based Access Control (RBAC) \cite{ferraiolo2001proposed}, Attribute-Based Access Control (ABAC) \cite{hu2013guide}, Data Usage Control (DUC) \cite{sandhu2003usage, park2004uconabc}, Group-Based Access Control (GBAC) \cite{GBAC}, Context-Aware Access Control (CAAC) \cite{toninelli2006semantic}, Risk-Aware Access Control (RAAC) \cite{atlam2020risk}, Relationship-Based Access Control (ReBAC) \cite{fong2011relationship}, and Trust-Based Access Control (TBAC) \cite{almenarez2005trustac, mahalle2013fuzzy}.  Besides, there are also typical cryptography-based access controls, such as Attribute-Based Encryption (ABE) \cite{huang2020attribute}.

Implementing these access control mechanisms confronts several significant challenges, which edge computing introduces \cite{xiao2019edge, khan2019edge}. First, a wide range of IoT devices in edge computing often operate with limited resources like storage, computational power, and communication. The resource-constrained environments challenge the effective implementation of access control mechanisms that typically require many resources. Second, the latency in edge computing results from various factors like task offloading, communication duration, computational expenses, and security measure-induced overheads. To preserve its low latency advantage, the implementation of access control systems must not significantly extend these delays.
Third, the mobile nature of data producers and consumers in edge computing scenarios necessitates flexibility in access control mechanisms to accommodate this dynamism. 
Finally, the widespread geographical distribution of edge nodes challenges the scalability of centralized access control systems and complicates trust management in such a decentralized environment.

Related surveys, from different viewpoints, have been published to review the existing access control measures in IoT edge computing, which attempt to conquer these challenges for achieving resource-conserving, low-latency, flexible, and scalable access control.
For instance, several surveys were reviewed from the point of different typical access control models \cite{kayes2020survey, alnefaie2021survey, ABESurvey}. Further, several were reviewed blockchain-based access control, which provides several benefits for access control in IoT edge computing (e.g., auditability, and scalability) \cite{pal2022blockchain, khan2022authorization, drame2021centralized}. Because IoT data usually has multiple lifecycles \cite{atzori2010internet},  several were reviewed from the point of different data lifecycles \cite{khalid2021survey, liu2019survey}.
However, the existing surveys have several limitations, which could hinder the development of access control in IoT edge computing.
Firstly, access control models are reviewed partly or nonsystematically. 
Secondly, although data lifecycles-based surveys can do a systematic review, they do not comprehensively consider those typical access control models, correspondingly some without the related blockchain-based schemes. Consequently, they also lack the corresponding systematic discussions and insights because of the first two limitations. These limitations motivate us to have a comprehensive and systematic survey with the following contributions: 

\begin{itemize}
    \item 
According to the end-edge-cloud architecture, we analyze the access control requirements in multiple data lifecycles in IoT edge computing, i.e., data collection, data storage, and data usage.
With the organized data lifecycles, we comprehensively show how to use/enhance the CapBAC, ABAC, ABE, DUC, RBAC, GBAC, CAAC (basic, TBAC, RAAC, and ReBAC) in IoT edge computing, and integrate blockchain for securing critical access control data.

 \item 
     According to involved data lifecycles in the systematic requirements analysis, we present and discuss the recent studies on access control in IoT edge computing, and provide the lessons learned from the involved studies. For example, with the organized data lifecycles in this survey, we find that the components used in typical access control models are not thoroughly explored for enhancing the adaptivity of access control in IoT edge computing. We summarize several rules for improving the efficiency and effectiveness of access control when using complex cryptography-based access control, e.g., reducing human power by machine learning. We observe that each data lifecycle has a limited number of blockchain-supported models, which means that the development of blockchain-based access control lags behind the innovative access control measures and thus cannot protect the corresponding critical access control data. The current blockchain-based versatile access control platforms, which intend to satisfy various access control requirements for multiple data lifecycles, mainly support conventional models (e.g., ABAC, RBAC, and CapBAC), with lag.

\item According to the discussions about the recent studies, we further discuss several challenges and research directions of access control in IoT edge computing. For example, we discuss and envisage the development of machine learning-based access control in IoT edge computing by expanding limited access control datasets, introducing federated learning for access control, and inspecting adversarial attacks. Some hybrid strategies can be explored to improve access control across multiple data lifecycles or multiple access control models in IoT edge computing. Moreover, we propose to study the powerful access control technology testbed based on the recent simulation platforms for IoT edge computing, to mitigate the complexity of systematically validating access control measures when various access control models are involved in multiple data lifecycles.
    
\end{itemize}

\begin{figure}
	\begin{center}
	\includegraphics[width=\columnwidth]{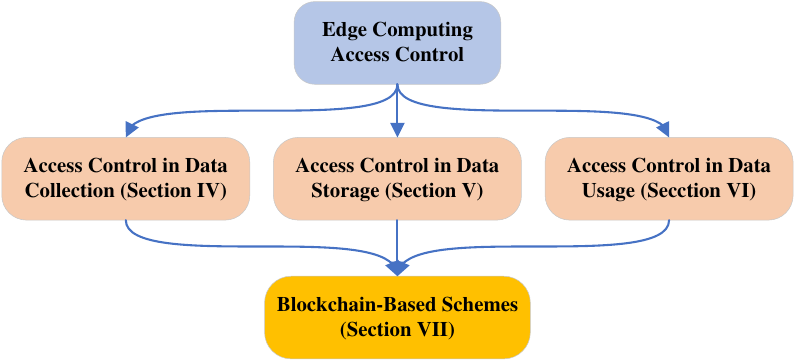}
		\caption{Classification of related studies to be discussed in this survey.}
		\label{Fig::Classification}
	\end{center}
\end{figure}

For the reader's convenience, we classify the recent studies to be discussed in this survey in Figure \ref{Fig::Classification}.
\textit{Our classification does not exhaustively present all kinds of data lifecycle, and access control models for each data lifecycle, due to the complexity of access control in IoT edge computing. However, this survey can complement existing surveys, which can help readers understand edge computing access control systematically. Further, the researcher or developer in the field can make systematical selections and innovations based on the related access control techniques. 
This survey can prompt the development of access control techniques in edge computing under the intricacy of innovations in access control.}


The rest of the paper is organized as follows. Section \ref{Sec::TraAccessControl} introduces typical access control models and blockchain technology with meaningful insights. Subsequently, Section \ref{Sec::RelatedSurveys} introduces related surveys.
Section \ref{Sec::RAoAC} describes background knowledge about IoT edge computing and then presents our requirements analysis on access control in IoT edge computing. 
Sections \ref{Sec::ACInDataCollection}\&\ref{Sec: ACInDataStorage}\&\ref{Sec::ACInDataUsage} respectively introduce the recent access control solutions according to the requirements. 
Further, Section \ref{Sec::Discussions} envisages some potential research directions to foster research efforts. 
Section \ref{Sec::SumAndCon} concludes this paper.  

\section{Typical Access Control Models and Blockchain Technology}
\label{Sec::TraAccessControl}
This section introduces some typical access control models and basic concepts in blockchain. This section can help readers understand the related surveys, our access control requirements analysis, and the reviewed studies in this survey.
\subsection{Typical access control models}
The access control model formally defines how subjects, objects, actions, conditions, and authorization policies are represented \cite{ferrari2010access}. And, the authorization policies regulate which object(s) can be accessed by which subject(s) with which action(s) under what condition(s). 
An access control mechanism is enforced according to an access control model and ensures that the objects can be controlled and used legally by reasonable access control.

Many typical and popular access control models are available in academia and industry, such as,
Capability-Based Access Control (CapBAC) \cite{hernandez2013distributed}, Role-Based Access Control (RBAC) \cite{ferraiolo2001proposed}, Attribute-Based Access Control (ABAC) \cite{hu2013guide}, Data Usage Control (DUC) \cite{sandhu2003usage, park2004uconabc}, Group-Based Access Control (GBAC) \cite{GBAC}, Context-Aware Access Control (CAAC) \cite{toninelli2006semantic}, Risk-Aware Access Control (RAAC) \cite{atlam2020risk}, Relationship-Based Access Control (ReBAC) \cite{fong2011relationship}, and Trust-Based Access Control (TBAC) \cite{almenarez2005trustac, mahalle2013fuzzy}. To organize this survey, we classify RAAC, ReBAC, and TBAC into CAAC, and we will explain in the following. 

Access Control Lists (ACLs) \cite{grunbacher2003posix} are the most common access control, where the access right on resources is represented by a sequence of pairs $\langle$subject, operations$\rangle$. However, the list would become complex with the increment of subjects and resources.
To overcome the deficiency, RBAC \cite{ferraiolo2001proposed} introduces the role concept, which decouples the subjects and resources by assigning rights to the role(s) of subjects. Thus, RBAC reduces the efforts to manage access rights. However, in a large number of resources or cross-organization scenarios, the role explosion still needs lots of effort to be conquered. ABAC \cite{hu2013guide} uses properties of subjects, objects (resources), and environment to specify access rights, which exploits more components to manage access rights. This fashion could mitigate the role explosion problem but with the expense of more complex attributes components and access rules. However, the common problem of ACLs, RBAC, and ABAC is that they enforce access control by checking the union of all rights in a request.  
To avoid unnecessary traversal in the checking, CapBAC \cite{hernandez2013distributed} authorizes access rights by subject-oriented linked list,
a sequence of pairs $\langle$object, rights$\rangle$. CapBAC could achieve minimal privilege enforcement by checking the request-specified legal capability. Efficient capability revocations and updating can be easily enforced. 
However, CapBAC needs to issue capabilities for all subjects by cryptographic techniques.

GBAC is a group-based model, where entities with the same permission requirements are assigned to the same group. Each group has specific access rights and attributes. Administrators can create groups of different levels as needed to achieve finer access control.
The inappropriate management of groups likely induces poor access control performance.

We can further incorporate context information in access control to enhance access permission management. 
In CAAC, access permissions are determined based on the combination of the subject, object, and other contextual information. CAAC can utilize this contextual information to adjust access permissions dynamically, thereby better meeting specific requirements and enhancing security. Contextual information may include basic ones, such as time, location, device status, network conditions, etc, and derived ones, such as trust, risk, relationship, etc. We mainly introduce the TBAC, RAAC, and ReBAC for this survey. 1) TBAC: In the era of pervasive computing resources and IoT devices, it is common that trust between entities cannot be established beforehand. TBAC focuses on enforcing access control by evaluating the trust level of the subject or object when trusted identities are not determined mutually. 2) RAAC:
In RAAC, the authorization of an access request depends 
on the risk factors associated with that request. These risk factors may include user behavior patterns, device security status, network security status, etc. The system dynamically adjusts access permissions based on these risk factors to provide appropriate resource access at different risk levels. 3) ReBAC:
ReBAC emphasizes the relationship between subjects in authorization decisions. The relationship can be diversified, such as superior-subordinate, cooperative, friendly, etc. Administrators can use the relationship to define access rights.

The above access control models are non-cryptographic access control, which only controls data access according to access control policies. However, the data can be easily leaked if the access control is bypassed. With cryptography for access control, we can control data access according to access control policies while the protected data is encrypted \cite{huang2020attribute}.
The Attribute-Based Encryption (ABE) is a popular scheme in this kind of approach.
In an ABE-based system, data decryption is contingent upon the alignment of attributes. Specifically, both ciphertexts and users’ keys carry specific attributes. A user can only decrypt the ciphertext if the user’s attributes (or access policy) match the ciphertext’s access policy (or attributes). 
Key-Policy Attribute-Based Encryption (KP-ABE) combines access policy with key and encrypts data with attributes. Ciphertext-Policy Attributes-Based Encryption (CP-ABE) combines attributes with keys and encrypts data with access policy.  
ABE-based access control has mobility, i.e., each distributed data-hosting party can be a policy enforcement point because the ciphertext of data is combined with the access control. However, as asymmetric encryption technology, ABE requires the extra management of the ciphertext and keys and is generally used for small data volumes.

\begin{figure}
	\begin{center}
	\includegraphics[width=\columnwidth]{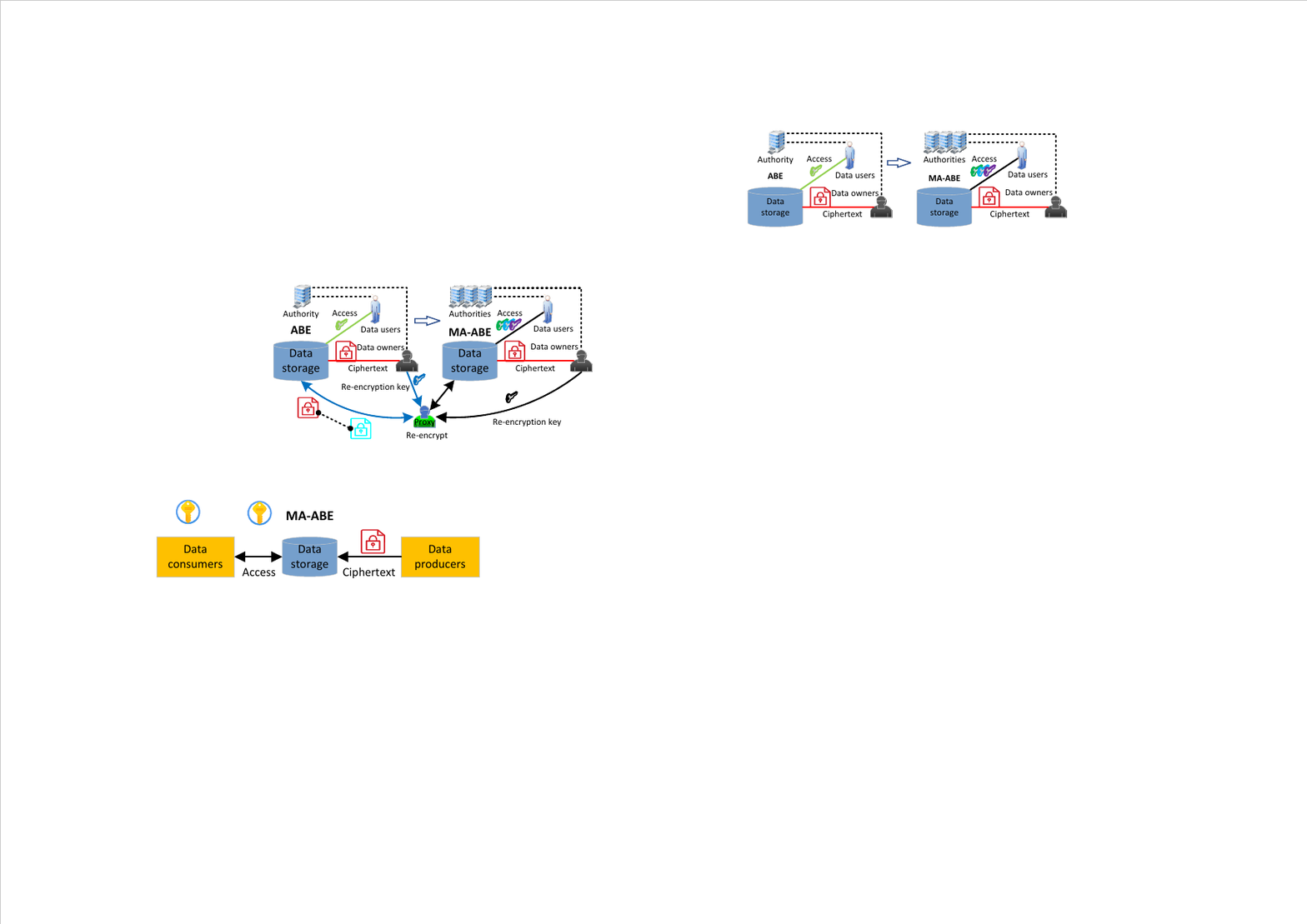}
		\caption{Attribute-based encryption (ABE) and multi-authority attribute-based encryption (MA-ABE) \cite{huang2020attribute}.}
		\label{Fig::ABEAndMAABE}
	\end{center}
\end{figure}

As shown in Figure \ref{Fig::ABEAndMAABE}, traditional ABE requires the authority to manage keys based on legal attributes. Data owners and data users only interact with the key intermediary. The abnormal key intermediary (e.g., disrupted legal attributes, DOS attack because of hacker intrusion) can affect the whole ABE system. Otherwise, multi-authority ABE (MA-ABE) depends on many related attribute authorities, where we can configure attributes used for encryption and decryption. For example, we can set that decryption can succeed with at least $d_{k}$ authorities providing secret keys for data users. As such, we can use the MA-ABE to manage attributes distributedly and resolve the key escrow problem.

The above models have a common drawback: they lose control after the data is granted access. DUC restricts data usage to enforce more flexible control after the data has been granted. For example, a party can use DUC to restrict data usage after it shares its data with another party. The restriction rules are specified by data owners who want to control how their data can be used. DUC can satisfy multiple data sovereignty requirements \cite{otto2022designing}. For example, the data usage scope controls that data must be used by trusted nodes or services; the data usage purpose controls that data must be used for specific usage purposes; the living time controls that data must be deleted immediately after data is used. However, DUC needs additional expenditure to accomplish these functions, such as purpose management and recognition.

\subsection{Blockchain technology}
\label{Sec::BlockchainTech}
The blockchain is a kind of distributed ledger technology \cite{dabbagh2021survey, zhang2021resource, zheng2018blockchain}. The committed transactions are recorded in the block. The linked block has the cryptographic hash summary of the content of the previous block, which enables the blockchain to be tamper-evident and tamper-resistant. Storing data in a distributed blockchain network reduces the risk of a single point of failure. The coded contract in a smart contract on a blockchain can be deployed to and executed by blockchain nodes. And, smart contracts can satisfy the common contract items and the need for trusted intermediaries. The execution results are ensured to be correct due to the consensus mechanism in blockchain so that the verification of calculation results can be avoided. 

Blockchain can be divided into permissionless and permissioned blockchains \cite{dabbagh2021survey}. Permissionless blockchain allows anyone to join, e.g., the popular public blockchain systems, bitcoin \cite{nakamoto2008bitcoin} and Ethereum \cite{wood2014ethereum}. Permissioned blockchain allows legal users of blockchain to participate, e.g., the consortium
blockchain, Hyperledger Fabric \cite{androulaki2018hyperledger}.

\section{Related Surveys}
\label{Sec::RelatedSurveys}

Kayes et al. \cite{kayes2020survey} make the review for CAAC. They systematically review the basic contextual conditions, e.g., temporal, location, and spatial, and the derived ones, e.g., relationship. And, they introduce how CAAC works with RBAC and ABAC for adapting to the dynamic environment in edge computing.
Zhang et al. \cite{zhang2018survey} present CapBAC, ABAC, ABE, RBAC, and DUC in edge computing. However, they do not cover GBAC and CAAC. 
Alnefaie et al. \cite{alnefaie2021survey} survey the CapBAC, ABAC, DUC, and RBAC, meanwhile, introducing CAAC (basic and TBAC), but without ABE, GBAC, RAAC, and ReBAC. Rasori et al. \cite{ABESurvey} focus on reviewing suitable ABE for IoT, which assumes that the edge servers are untrusted. To spur the development of DUC, Akaichi et al. \cite{akaichi2022usage} review various DUC proposals for their specification, enforcement, and robustness.

While the aforementioned studies make their reviews with different typical access control models, they do not consider blockchain-based access control, which is an emerging novel access control technology.
Pal et al. \cite{pal2022blockchain} mainly present the studies to use blockchain for enhancing CapBAC, RBAC, TBAC, and ABAC in edge computing. 
Riabi et al. \cite{riabi2019survey} present blockchain-based access control for improving ABAC, RBAC, and CapBAC.
Khan et al. \cite{khan2022authorization} classify the existing authorization mechanisms, weaknesses, and loopholes. 
The CapBAC, ABAC, ABE, TBAC, and their corresponding blockchain-based schemes are involved. 
Dram{\'e}-Maign{\'e} et al. \cite{drame2021centralized} survey the access control solutions from the perspectives of architectures, e.g., centralized, hierarchical, federated, and distributed. The CapABC, ABAC, ABE, RBAC, CAAC, and general blockchain-based access control are analyzed in these architectures.

Moreover, 
Ravidas et al. \cite{ravidas2019access} divide the requirements of access control systems into three categories according to policy lifecycles, i.e., policy specification, policy management, and policy evaluation\&enforcement. This survey analyzes CapBAC, ABAC, DUC, RBAC, and CAAC according to the above classifications.

However, the aforementioned studies do not consider the reviews from the perspective of multiple data lifecycles. In edge computing, data security encompasses multiple stages including data collection, storage, usage, etc. A comprehensive review of access control techniques across multiple data lifecycles can facilitate the holistic technique advancement.
Khalid et al. \cite{khalid2021survey} survey on secure data storage and retrieval in edge computing. 
It involves ABE and TBAC and discusses the corresponding management issues.
Liu et al. \cite{liu2019survey} conduct a comprehensive review on secure outsourced data analytics in edge computing. 
Secure data collection covers authentication mechanisms and trust management mechanisms. 
Secure data processing relies on the following techniques: homomorphic encryption, differential privacy, pseudonym technology, etc. 
Data processing generates computational results, which could be stored and kept protected. 
Secure data storage usually leverages access control mechanisms (e.g., ABE and RBAC) and searchable encryption (including symmetric and asymmetric methods). However, this work does not involve CapBAC, DUC, GBAC, and the related blockchain-based schemes.

\section{Edge Computing and Requirements Analysis on Access Control}
\label{Sec::RAoAC}

\subsection{IoT edge computing}
\label{Sec::Background_EdgeComputing}

Traditionally, the central cloud is the main platform to provide computing power for a wealth of data from distributed heterogeneous IoT devices \cite{botta2016integration}. 
The traditional cloud-centric architecture involves long propagation delays, which is unacceptable for latency-sensitive applications \cite{lim2020federated}. 
We can find some latency-sensitive examples in \cite{varghese2016challenges}. For example, in a smart health system, a large amount of data generated by smart sensors for telemedicine needs to be processed in microsecond latency. 
In the telemedicine scenario, the high-solution videos need to be shown to doctors in time.
A visual guiding service using a wearable camera requires a preferred response time between 25 ms to 50 ms.

\begin{figure}
	\begin{center}
	\includegraphics[width=\columnwidth]{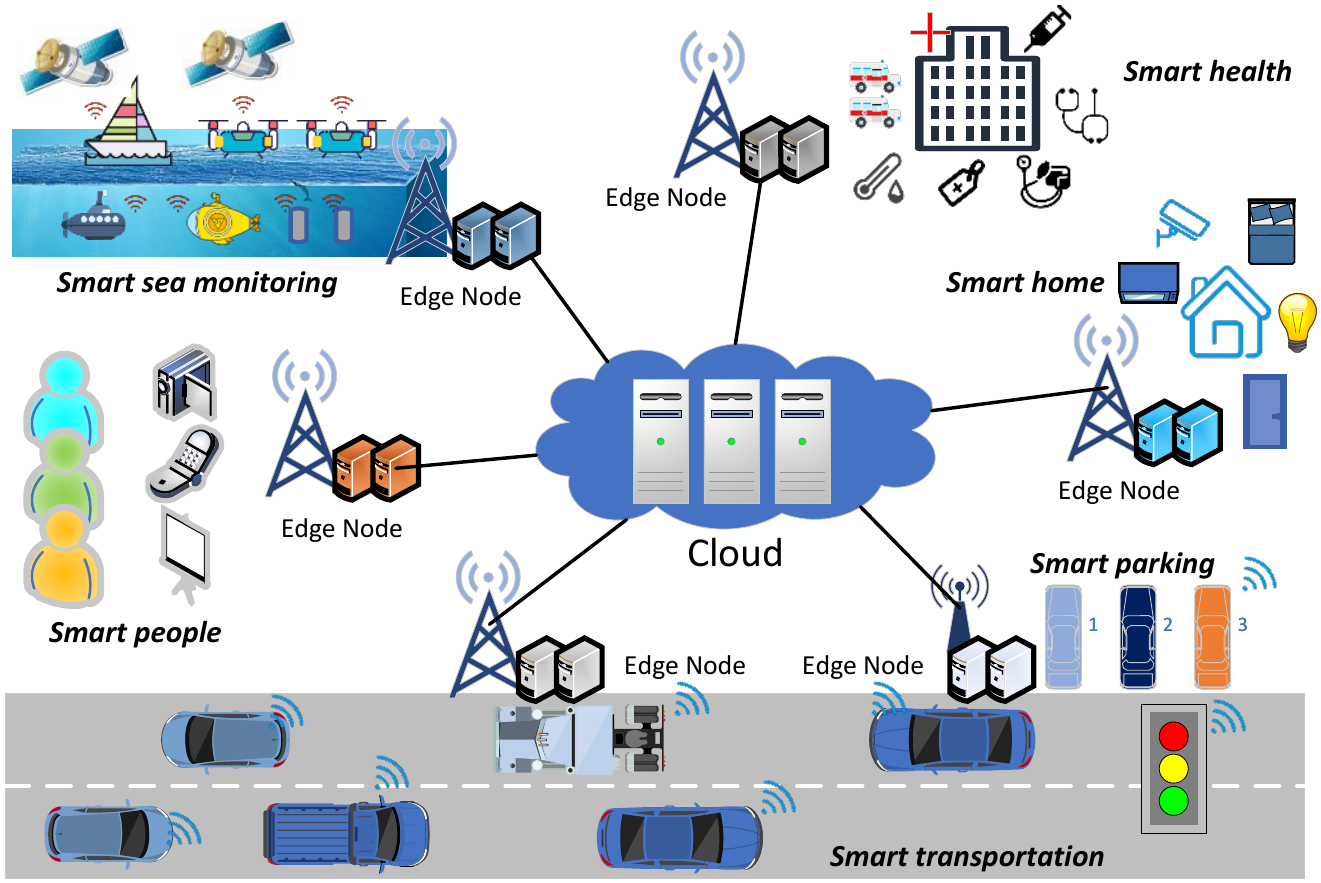}
		\caption{The end-edge-cloud system architecture \cite{shi2016edge}.}
		\label{Fig::EdgeComputing}
	\end{center}
\end{figure}

To conquer the latency challenges in traditional cloud computing, the end-edge-cloud system architecture, as shown in Figure \ref{Fig::EdgeComputing}, is proposed with these layered and connected components: cloud, edge nodes, and end devices \cite{shi2016edge}, where data can be processed closer to the data source. 
In particular, edge nodes connect with pervasive heterogeneous IoT devices, and interact with the cloud, meanwhile, an edge computing orchestrator could orchestrate cooperative work among edge nodes/cloud \cite{ren2019survey}. 
Massive data collected from IoT devices can be aggregated at edge nodes distributedly before the data is delivered to the cloud. Real-time data processing and basic analytics can be fulfilled at the edge.
Due to the advantages of edge computing, it has been widely used in many applications as shown in Figure \ref{Fig::EdgeComputing}, including smart transportation, smart sea monitoring, smart homes, smart health, etc.

\subsection{Requirements analysis on access control}

This section would describe the access control requirements for edge computing in the tri-layer end-edge-cloud architecture.
Traditional cloud computing requires large raw data to be centrally uploaded to the cloud servers for unified processing, which increases the risk of data loss and leakage \cite{cao2020overview}. 
However, IoT edge computing enables data to be processed locally by deployed edge servers, reducing the need to transfer data to the cloud servers for processing, thus enhancing data privacy and security. 
With massive data processed at the edge, access control, as the common fundamental technique for protecting data, should be studied thoroughly. 
To have a clear analysis of access control requirements, we first show the following natural characteristics of IoT edge computing, which would challenge the access control at the edge \cite{xiao2019edge, khan2019edge}:
\begin{itemize}
\item 
\textit{Limited resources:} As shown in Figure \ref{Fig::EdgeComputing}, various smart devices are at the end. However, these devices often have constraints on resources (e.g., energy, storage, computation, and communication), while executing access control generally requires enough resources.

\item 
\textit{Low latency:} Edge computing embodies latency-sensitive applications. In particular, the latency in edge computing includes many aspects: task-offloading time, communication latency, computation cost, security-induced cost, etc. \cite{liu2019dynamic, kai2020collaborative}.
The total of these delays should be as small as possible.
The low latency of edge computing should not be negated by those aspects. 

\item
\textit{Flexibility:}
Data producers and consumers are generally dynamic. In particular, unbound IoT devices may be added; short-lived interaction is typical; services are potentially changing \cite{khan2019edge}. As such, the dynamic nature of data producers and consumers in edge-computing scenarios necessitates flexibility in security measures to accommodate this dynamism.

\item
\textit{Scalability:} The decentralized architecture of edge computing allocates computing resources at scale  \cite{khan2019edge}. In this way, the widespread geographical distribution of edge nodes would naturally challenge the scalability of centralized security mechanisms and complicate trust management in such a decentralized environment.

\end{itemize}

\begin{figure}
	\begin{center}	\includegraphics[width=\columnwidth]{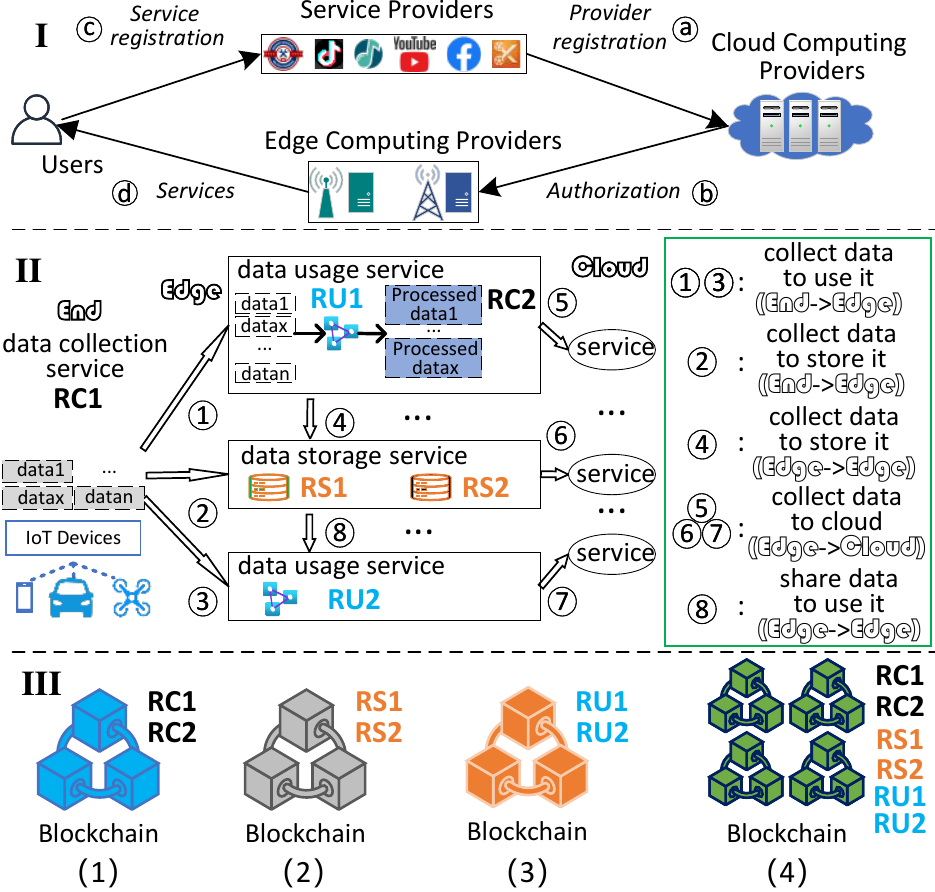}
		\caption{Access control requirements analysis in edge computing.}
		\label{Fig::DataACRE}
	\end{center}
\end{figure}

Corresponding to the above characteristics of IoT edge computing, the access control of edge computing should conserve resources and have low delay, flexibility, and scalability. The typical models and blockchain technology introduced in Section \ref{Sec::TraAccessControl} can be developed to satisfy the requirements. Specifically, we would analyze how to satisfy these requirements in this survey from the perspectives of the following common data lifecycles: data collection, data storage, and data usage \cite{DLM}.

To clarify the requirements analysis, we first generalize the basic process of service provisioning and utilization in the tri-layer end-edge-cloud architecture according to the studies \cite{wang2020convergence, fan2021serving, bozorgchenani2018centralized}. 
Figure \ref{Fig::DataACRE} (I) shows four kinds of stakeholders in edge computing services: users, service providers, edge computing providers, and cloud computing providers \cite{ning2020distributed}. 
Generally, service providers register on the cloud  (\textcircled{a}), and cloud computing providers authorize corresponding services on legal edge servers (\textcircled{b}). 
Then, users register services from service providers(\textcircled{c}), and users can get registered services from edge servers of edge computing providers (\textcircled{d}).
In this paper, IoT devices and edge services act as various users. 
To describe the access control requirements, we generalize three kinds of services here, i.e., data collection service, data storage service, and data usage service (as shown in Figure \ref{Fig::DataACRE} (II)), which correspond to each data lifecycle respectively.

\textit{Data Collection:} 
As shown in Figure \ref{Fig::DataACRE} (II), IoT devices generate large amounts of data. 
Because IoT devices commonly have limited computation and storage resources or some services require various data from IoT devices, IoT-generated data is uploaded to edge servers for the related services. In particular, as shown in the green box in Figure \ref{Fig::DataACRE} (II), \ding{172}\ding{174}arrows indicate the end$\rightarrow$edge data collection, i.e., collecting data to use it at edge; \ding{173} arrow indicates another end$\rightarrow$edge data collection, i.e., collecting data to store it at edge.
In the process, access control should determine which IoT-generated raw data can be accessed by which services, meanwhile the access control should be flexible given the flexibility of end devices and services (denoted as \textbf{RC1} requirement). 
The following access control models can be used for satisfying \textbf{RC1}: GBAC, ABAC, CapBAC, and CAAC. Therein, GBAC can facilitate data management by leveraging the group concept on IoT devices. 
ABAC could be suitable because it can adapt the across-domain data collection.
CapBAC consumes small resources of IoT devices due to its method of authorization of access rights. CAAC can use basic context and derived context information to further accommodate the flexibility of edge computing.

Due to the end-edge-cloud architecture, as shown in Figure \ref{Fig::DataACRE} (II), data collection from edge to cloud or from edge to edge is commonplace \cite{wang2020edge}. In particular, as shown in the green box in Figure \ref{Fig::DataACRE} (II), \ding{175} arrow indicates the edge$\rightarrow$edge data collection, i.e., some data usage services produce intermediate aggregation results, and data storage services may collect them to store them. 
\ding{176}\ding{177}\ding{178} arrows indicate the  edge$\rightarrow$cloud data collection, i.e., edge computing could achieve real-time data processing while cloud services collect/sample some data at the edge for further advanced analysis.
In this situation, access control should determine which intermediate data can be accessed by which edge services
or cloud services (denoted as \textbf{RC2} requirement).
The required access control techniques include CapBAC, ABAC, and CAAC similar to in \textbf{RC1}. Besides, ABE can prevent the collected small raw data from being leaked on malicious edge servers.

\textit{Data Storage:} Second, we inspect the access control for securely storing the IoT-generated data and its related computation results, so that the stored data can be securely used for mining valuable insights or interacting with the IoT devices.
In particular, due to limited resources on IoT devices, IoT-generated data would be stored on edge servers, as mentioned \ding{173} arrow; and the computation results of IoT-generated data could be stored on edge servers, as mentioned \ding{175} arrow. 
The stored IoT-generated data should be accessed by corresponding authorized parties, which have different access rights to data (denoted as \textbf{RS1} requirement). 
The stored computation results may cooperate among edge servers, where different edge services should access different computation results (denoted as \textbf{RS2} requirement).

\begin{table*}[]
\centering
\fontsize{10pt}{11pt}\selectfont{
\begin{tabular}{|l|c|c|l|l|c|c|c|}
\hline
\multirow{2}{*}{\diagbox{LC}{AC}} & \multirow{2}{*}{CapBAC} & \multirow{2}{*}{RBAC} & \multicolumn{1}{c|}{\multirow{2}{*}{ABAC}}& \multicolumn{1}{c|}{\multirow{2}{*}{ABE}} & \multicolumn{1}{c|}{\multirow{2}{*}{DUC}}& \multirow{2}{*}{GBAC} & \multirow{2}{*}{CAAC$_{a}$}  \\ 
                       &                                                &                       &                      & \multicolumn{1}{c|}{}                                      &                       &                       &  
                       \\ \cline{1-8} \xrowht{7pt}
DC                     &  \checked & - & \checked & \checked & -  &  \checked & \checked,\checked,\checked,-     \\ 
\cline{1-8} \xrowht{7pt}
DS      &   -    & \checked  &  \checked & \checked &  - &  -                  &   \checked,\checked,\checked,-     \\ \cline{1-8} \xrowht{7pt}
DU    &   -    &  \checked  &  -   &  -   &  \checked  & - &  \checked,\checked,\checked,\checked                                                                                                                                                    \\ \hline 
\end{tabular}

\vspace{1mm}
   \begin{tablenotes}
     \item[1] AC: Access Control, LC: Lifecycle, DC: Data Collection, DS: Data Storage, DU: Data Usage, (CAAC$_{a}$=(basic, TBAC, RAAC, ReBAC)
   \end{tablenotes}
}
\caption{Involved typical access control models in each classification in this survey.}\label{Table::ACReuqirementsTable}
\end{table*}

Data cryptography for access control is practical for satisfying \textbf{RS1} \& \textbf{RS2}.
Edge servers commonly belong to third parties.
The encrypted data ensures that malicious third-party edge servers cannot peep data easily, meanwhile, the native access control functions by cryptography can control which subject can access which part of encrypted data \cite{di2007data}.
ABE-based access control is promising, where various attributes, e.g., role, context, trust, risk, etc. can be used for regulating access policies flexibly.
Moreover, ABE-based access control has mobility, i.e., each distributed data-hosting party can be a policy enforcement point because the ciphertext of data is combined with the access control. 
It is a data-centric access control without centrally depending on a policy enforcement point. 
ABE could prompt data sharing with high flexibility and scalability in edge computing.

\textit{Data Usage:} As shown in Figure 
\ref{Fig::DataACRE} (II), the edge computing ecosystem has pervasive data analytics. In particular, data analytics exists in some edge servers, whose responsibility is to analyze raw data. For instance, as mentioned \ding{172}\ding{174} arrows, the collected raw data is to be used. Besides, data analytics also exists in some edge servers, which use some computation results shared from other servers for high-level analysis, as \ding{179} arrow indicates.
Data owners could expect to learn how their outsourced data is used and control its usage (denoted as \textbf{RU1} \& \textbf{RU2} requirements).
However, the above access control models in data collection and storage would end their responsibility for protecting data after data access is granted. DUC could satisfy data sovereignty requirements because it enforces further flexible control by data usage restrictions.

\textit{Securing Critical Access Control Data:} In summary, because the typical access control models mentioned in the above data lifecycles are not derived from edge computing natively, we should attempt to enforce them effectively in distributed edge computing to achieve low resource consumption and delay for access control, while ensuring flexibility and scalability. However, an access control system generally includes a wealth of sensitive access control data \cite{BlockchainforAC}.
For example, CapBAC includes authorization tokens, token registration, delegation, and revocation functions.
ABAC contains various attributes, attribute-based policies, and programs for enforcing policy. Some important access control data should be protected, e.g., sensitive tokens, private attributes, critical policies, and enforcement programs.

Blockchain can be used to store important access control data because, as introduced in Section \ref{Sec::BlockchainTech}, it is a distributed database for storing data and the on-chain data is immutable. Moreover, accountability is natively provided because blockchain has immutability and traceability properties. 
Besides, smart contracts can be coded distributedly to enforce access control and the security of enforcement results is guaranteed by the consensus mechanism, and it can be used as a trustable alternative for enforcement programs. 
As shown in Figure \ref{Fig::DataACRE} (III)(1)(2)(3), blockchain technology can enhance the access control mechanisms in data collection, storage, and usage, respectively. More precisely, the access control requirements for each data lifecycle can be achieved while safeguarding critical access control data.

Furthermore, the blockchain platform can also manage various access control enforcement. In particular, the blockchain platform can embody multiple access control functions, e.g., CapBAC for data collection, ABE for data storage, and DUC for data usage.  
That is, as shown in Figure \ref{Fig::DataACRE} (III)(4), a platform could satisfy all access control requirements (\textbf{RC1}\&\textbf{RC2}, \textbf{RS1}\&\textbf{RS2}, and \textbf{RU1}\&\textbf{RU2}) meanwhile access control data for multiple data lifecycles is uniformly managed.

\textit{Summary:} We have described the access control requirements by the common multiple lifecycles (including data collection, storage, and usage) while analyzing blockchain's role in protecting critical access control data. As shown in Table \ref{Table::ACReuqirementsTable}, we summarize the adopted typical access control models for satisfying the access control requirements in each data lifecycle.
The contextual properties of CAAC can be integrated into ABAC, ABE, and DUC. In particular, the environmental properties of ABAC can use the contextual properties of CAAC to achieve access control; ABE can use them to encrypt/decrypt sensitive data; DUC can use them to control how to use data. Moreover, the role in RBAC can be regarded as an attribute of ABAC/ABE and a component of DUC.
The integration can facilitate the flexibility of access control. 
Finally, because consensus algorithms in blockchain possibly induce large latency, we should protect critical access control data with acceptable latency.

In the following sections, we will describe the recent solutions to use these typical access control models as base models for implementing access control in edge computing: 1) access control in data collection (Section \ref{Sec::ACInDataCollection}), 2) access control in data storage (Section \ref{Sec: ACInDataStorage}), 3) access control in data usage (Section \ref{Sec::ACInDataUsage}), and 4) blockchain-based access control platform (Section \ref{Sec::BBACP}).


\section{Access Control in Data Collection}
\label{Sec::ACInDataCollection}

An abundance of raw data measured from IoT devices should be securely collected for subsequent processing and valuable insights. 
Access control should determine which IoT-generated raw data can be accessed by which services, meanwhile the access control should be flexible given the flexibility of end devices and services. For the processed raw data (i.e., intermediate data), access control should determine which intermediate data can be accessed by which edge service or cloud service.
In this section, we present the studies according to the typical model used as the base model to meet the access requirements, with the summary and lessons learned from these studies. 

\textit{GBAC as the base model:} To protect IoT-generated data from tampering and unauthorized access, group key management (GKM) is proposed \cite{grammatikis2019securing}. A device group key is shared with its current subscribers, and the device can encrypt its data by using a group key and only the legal subscribers can decrypt it. This ensures that authorized group users can access the data collected from IoT devices. However, the GKM only assigns a symmetric group key to each group and does not support members' independence in the same group. This means that keys need to be reassigned every time membership changes or state changes in a group. The flexibility in the IoT environment can induce frequent key redistribution and thus introduce large management and communication overhead. To conquer the challenges, Dammak et al \cite{dammak2020decentralized} propose a new decentralized lightweight group key management architecture for access control. It introduces a layered architecture, where a central key distribution center accounts for the management of the device group. 
Each device group has several user groups, and each user group corresponds to a sub-key distribution center (SKDC). By dispersing the key management tasks across multiple SKDCs, the system can better handle the complexity and overhead caused by changes in membership. When a user joins or leaves a group, only the SKDC where the user is located is adjusted.

\textit{TBAC as the base model: }Jiang et al. \cite{jiang2022trust} present a trust-based energy-efficient data collection for unmanned aerial vehicles (UAV). A UAV trajectory optimization algorithm recognizes the most data collection anchors, which guarantees the data collection trajectory is shortest and extends network life, meanwhile, a trust reasoning and evolution mechanism is used to identify the trusted sensor nodes and guarantee the quality of collected data. This method increases the network life by 48.9\%, and achieves 91\% accuracy in node trust degree identification while consuming only 8\% of network life.

\textit{RAAC as the base model:} Atlam et al.\cite{atlam2021fuzzy} propose a dynamic and adaptive RAAC model. 
The risk assessment algorithm considers contextual information, resource sensitivity, action severity, and previous risk history. This model can adapt to the dynamics of IoT. When an access request is issued for data collection, the model can use the real-time characteristics of user requests to estimate the security risk and make an access decision. However, this solution is not evaluated.


\begin{table*}[]
\centering
\fontsize{10pt}{11pt}\selectfont{
\begin{tabular}{|c|p{1.5cm}|c|p{3.2cm}|p{3.2cm}|}
\hline
Ref & Base Model & Requirements & Performance & Key Improvements \\ \hline 
\cite{grammatikis2019securing, dammak2020decentralized}
   & GBAC  & \textbf{RC1}     &  High flexibility and scalability, Very low storage costs. Both computation and communication costs are microseconds.  
   & Introducing a decentralized lightweight group key management scheme for flexible and scalable data collection.                            \\ \hline 
\cite{jiang2022trust}
   & TBAC  & \textbf{RC1}       &  Increasing the network life by 48.9\%, and achieving 91\% accuracy in node trust degree identification while consuming only 8\% of network life.
   & Significantly improving the identification accuracy of node trust degree and greatly prolonging the network life.                            \\ \hline 
 \cite{atlam2021fuzzy} 
    & RAAC  & \textbf{RC1}      & -        
    & Attempting to provide the algorithm for calculating the risk of user requests to adapt the flexibility and scalability of IoT devices. \\ \hline
\cite{dougherty2021apecs, AADEC, liu2022secure}
    &  CapBAC    & \textbf{RC1} \& \textbf{RC2}    &  User token verification (4.8ms average) is efficient for IoT devices and edge servers.  & Providing a distributed access control framework for pervasive edge computing services.    \\ \hline

     \cite{alrawais2017attribute}
    & CP-ABE   &  \textbf{RC2}       &    The time overhead involves the key generation, encryption, and decryption, with proportional to the number of attributes.                               & Providing a protocol for encrypted key exchange based on CP-ABE to establish secure communication between edge nodes and the cloud.                             \\ \hline  \cite{zhang2022achieving}
    & CP-ABE  &  \textbf{RC2}  & Bilateral fuzzy matching causes slightly increased encryption time. Meanwhile, outsourcing matching workloads to edge nodes decreases decryption and communication overheads significantly.
    & Achieving data sharing with bilateral fuzzy matching in a secure cloud-edge communication environment.                               \\ \hline 
\end{tabular}
}

\caption{Access control schemes in data collection.} \label{Table::ACCollection}
\end{table*}

\textit{CapBAC as the base model:} An advanced access control framework (APECS) for pervasive edge computing is proposed \cite{dougherty2021apecs}, where the legitimated end devices/edge servers can lightly obtain services permission of available edge servers, i.e., the access control for collecting data from end to edge and edge to edge is supported simultaneously. APECS framework depends on the CapBAC, where the authorized token for users by service providers is a tuple, including the unique identifier of the service provider, a list of service identifiers, a certificate of the user, a list of authorization levels, and the token's expiry time. The authentication and authorization tasks are directly delegated to semi-trusted edge servers, rather than centralized cloud servers, such that network communication beyond the edge is avoided. Moreover, APECS supports efficient and quick token revocation for users and edge servers. 
However, the authentication token contains the user's identity, so privacy leakage could happen once the user's identity is parsed by servers. An anonymous and auditable distributed access control framework for edge computing (AADEC) is proposed \cite{AADEC} for overcoming the deficiency. In AADEC, the user identity information is hidden, which raises the bar for attackers sniffing user privacy.
Nevertheless, because of the data confidentiality, it is likely to induce fake data spreading issues. AADEC allows a trusted third party to achieve malicious data auditing or tracing, which further mitigates the threat in edge computing. Moreover, Liu et al. \cite{liu2022secure} improve the efficiency of access control in APECS by achieving batch service authorization.


\textit{CP-ABE as the base model:} 
To guarantee the confidentiality and verifiability of transmitted encrypted data between edge nodes and the cloud,
ARWA et al. \cite{alrawais2017attribute} design an encrypted key exchange protocol based on CP-ABE and digital signature. In particular, the edge node's shared key can only be obtained by the cloud or other edge node(s) with a set of legal attributes. The experimental results show that, in terms of the CP-ABE mechanism, the key generation time, the encryption time, and the decryption time are increased linearly with the number of attributes.
However, the attributes in this protocol are defined centrally, limiting access control's flexibility.
To implement more flexible access control, Zhang et al. \cite{zhang2022achieving} propose a scheme FADS, where edge nodes and cloud can specify access policies respectively and fuzzy matching is allowed. 
The transmitted data can be decrypted if and only if the receivers' attributes successfully match the access policies defined by two sides (sender and receiver). This matching is fuzzy, that is, receiver attributes can satisfy access policy without a rigid access structure. This way can hide the accurate user attributes, and enhance privacy protection. 

\subsection{Summary and lessons learned}
\label{Sec::SALInDataCollection}

In this section, we have discussed the studies to satisfy \textbf{RC1 \& RC2}.
The approaches are summarized in Table \ref{Table::ACCollection}, and the learned lessons are as follows:

\begin{itemize}
    \item 
We can observe that these studies leverage various components to achieve access control, which facilitates data collection within the dynamic environment of edge computing. Some focus on developing algorithms to evaluate access control components, such as group, trust, and risk. 
Others, addressing the challenges of resource-constrained IoT devices, incorporate different elements to create valid tokens in CapBAC. The high-performance implementation of CapBAC enables the effective use of edge services for processing IoT raw data.
However, privacy concerns, particularly regarding the sensitive elements in tokens, must be also addressed \cite{AADEC}.
Furthermore, when combined with data cryptography, these access control schemes can ensure secure and flexible data collection for smart IoT devices \cite{dammak2020decentralized}. 

\item 
Although components such as group, trust, risk, raw context, capability, and cryptography facilitate adaptive data collection in the dynamic environment of edge computing, their integration has not been thoroughly explored. 
For instance, some IoT devices may integrate data from various sources to deliver specific services. A high-performance CapBAC scheme is desirable, which would support flexible and efficient token management—enabling capability registration, delegation, and revocation. It should also allow for federated operations, enabling collaborative service access across different data sources, and optimize service location using additional information, such as network topology from software-defined networks \cite{goransson2016software}, group dynamics, and trust levels.

\end{itemize}

\section{Access Control in Data Storage}
\label{Sec: ACInDataStorage}

As learned from typical access control models, ABE-based access control is popular data cryptography solutions for secure data storage in edge computing. The stored raw data should be protected by access control policies specified by data owners or administrators. The stored computation results may be cooperated among edge nodes/cloud, where different edge/cloud services should access different computation results.
This section introduces the corresponding recent studies. 

\subsection{ABE-based access control}
\label{Sec::Abe_based_AC}

We recommend the surveys \cite{ABESurvey, oberko2022survey} for readers to learn more about ABE.
We present the most recent studies, which include general ABE-based solutions, generalizing the common problems in edge computing, and domain-specific ones, which design ABE for edge computing of some domains. We first introduce the general ABE-based solutions.


The linear relationship between the ciphertext/secret key sizes and the number of attributes impedes the development of ABE-based solutions. 
To address this issue, Sarma et al. \cite{sarma2022macfi} propose a multi-authority CP-ABE scheme MACFI, which reduces the data size and outsources expensive decryption computation to edge nodes. 
MACFI achieves that the size of the secret key is kept irrespective of the number of attributes and that the ciphertext size is agnostic for the number of attributes and has a linear relationship with the number of authorities. 
MACFI further increases the efficiency of CP-ABE-based access control for edge computing in IoT. Moreover, each attribute authority in MACFI verifies a user, and allocates attributes based on the user's role (RBAC).
Xu et al. \cite{xu2020match} propose a new matchmaking attribute-based encryption (MABE) for enhancing the performance of KP-ABE. On the one hand, MABE uses collision-resistant hash functions to generate the ciphertexts, which decreases the encryption time for IoT devices.
On the other hand, like outsourcing ciphertext decryption to edge nodes, MABE also outsources heavy ciphertext identification to edge nodes, but edge nodes can leverage the access structures from data receivers to filter useless data in the ciphertext. As such, the decryption time is decreased largely. 
As shown in the experiments, this way not only decreases the computation overhead for IoT devices garnering useful data but also saves the cost of bandwidth. 

Changing the mathematics in cryptography is another way to cut down the computation burden. 
To consume much fewer resources, Rui et al. \cite{cheng2021efficient} propose a novel CP-ABE scheme based on elliptic curve cryptography. As a result, the expensive bilinear pairing operation in native CP-ABE is replaced with simple scalar multiplication. In addition, this study considers the time and location attributes for encryption and also outsources expensive computational tasks to edge servers. As shown by the experiments, the encryption of time and location attributes consumes a few scalar multiplication operations, and the decryption time has no obvious linear relationship with the number of attributes because the resource-limited IoT devices only calculate the scalar multiplication operation one time. 

The aforementioned studies commonly improve ABE-based access control efficiency by offloading the decryption computation burden to edge nodes.
To further improve the efficiency, Xu et al. \cite{xu2021server} reduce the cost of data decryption simultaneously by outsourcing the workloads from these aspects: data source identification, revocable storage, and decryption mechanism.

IoT gateway can sense the context information with IoT devices or vicinity users, e.g., location, time, and sensitivity. The context information often changes continuously and frequently, making the static and traditional security solutions inefficient. For example, Rui et al. \cite{cheng2021efficient} simply combine contextual attributes (time and location) with their CP-ABE solution. Arfaoui et al. \cite{CAABAC} propose a context-aware attribute-based access control (CAABAC) approach for providing data consumers with secure and adaptive remote control of smart IoT. CAABAC incorporates the CP-ABE with contextual information to achieve adaptive contextual data collection for data consumers, where contextual tokens are integrated into the CP-ABE's access structure. 
When data consumers decrypt the collected data, they are required an attribute set and context information to collectively satisfy the access structure. In this solution, IoT gateways act as edge nodes, which store the encrypted data. The experiments show that small computation cost is consumed on IoT devices.
However, the most existing CP-ABE solutions force data users to manually provide the contextual information for encrypting or decrypting data. This way indicates that malicious data users can conjecture which attributes are incorrect, such that it is possible to raise unauthorized access to information.
Ghosh et al. \cite{ghosh2021case} propose a context-aware attribute learning scheme CASE for tackling the issues. Depending on the edge intelligence, CASE can learn users' contextual attributes and generate attributes automatically, which could decrease manual operations and thus enhance the system's security. Meanwhile, it reduces the size of post-encryption data based on learned attributes, which increases the scalability of access control as discussed in the experiments. 
Because a trust degree has regional characteristics in edge computing,
Zheng et al. \cite{zheng2022adaptive} propose an access control scheme based on CP-ABE, which combines users' trust degree.
The scheme leverages the classification and recommendation algorithms for adaptively determining users' trust degree in different regions. 
Therein, the trust degree is divided into three dimensions: the basic trust degree, the behavior trust degree, and the recommendation trust degree. The number of attributes in the scheme is constant, which decreases the computational overhead.

\begin{table}[]
\centering
\vspace{-1cm}
\fontsize{10pt}{11pt}\selectfont{
\begin{tabular}{|c|p{0.8cm}|c|p{4.3cm}|p{4.3cm}|}
\hline
Ref & Base Model & Requirements & Performance & Key Improvements \\ \hline 
\cite{sarma2022macfi}
   & CP-ABE  & \textbf{RS1}\&\textbf{RS2}     & Computation cost: Key generation ($2N_{A_{u}}$+$\left|\mathbb{A}\right|$)$T_{G}$, Encryption ($3N_{A_{u}}$+2(n-$\left|\mathbb{P}\right|$))$T_{G}$ + $T_{G_{T}}$, Decryption ($\left|\mathbb{A}\right|$-$\left|\mathbb{P}\right|$)$T_{G}$+$N_{A_{p}}$$T_{G_{T}}$+$3N_{A_{p}}$$T_{p}$. Ciphertext size: $3N_{A_{p}}$$\left|G\right|$+2$\left|G_{T}\right|$, Key size: $\left|Z_{p}\right|$.
   & The size of the secret key is irrespective of the number of attributes and the ciphertext size 
has a linear relationship with the number of
authorities                           \\ \hline 
\cite{xu2020match}
   & KP-ABE  & \textbf{RS1}\&\textbf{RS2}       &  Computational complexity: Encryption $\mathcal{O}$($\mathcal{S}$+$\mathcal{R}$), Decryption $\mathcal{O}$($\mathbb{R}$). Space Complexity: Encryption key $\mathcal{O}$($\mathcal{S}$), Decryption key $\mathcal{O}$($\mathbb{R}$).
   & Filtering useless data in the ciphertext for IoT devices (as data receivers) according to more constraints of data producers and consumers.                            \\ \hline 
 \cite{cheng2021efficient}
    & CP-ABE  & \textbf{RS1}\&\textbf{RS2}       & 
    Computation cost: Encryption (4$N_{r}$+1)G, Decryption [at edge] ($D_{a}$+1)G, Decryption [at end device] G. Ciphertext size (3$N_{r}$+1)$|G^{'}|$, Public key size (2n+2)$|G^{'}|$, Private key size $|\mathbb{P}|$ $|G^{'}|$.
    & Cutting down the computation burden for resource-limited IoT devices by changing the mathematics in cryptography. \\ \hline
 \cite{xu2021server}
    &  CP-ABE    & \textbf{RS1}\&\textbf{RS2}   &  Computational complexity: Key generation $\mathcal{O}$(1), Encryption $\mathcal{O}$($\mathcal{S}$+$\mathbb{R}$), Decryption $\mathcal{O}$($1$). & Reducing the computation burden for resource-limited IoT devices from multiple aspects.   \\ \hline
    
    \cite{CAABAC}
    & CP-ABE  &   \textbf{RS1}\&\textbf{RS2}   &     Storage overhead: $(\left|\mathbb{A}\right|+2)\left|{G}_{1}\right|$
    ,
    Communication overhead: 359 bytes (IoT gateway), 16 bytes (Smart thing), Computation cost: 673.4ms (IoT gateway), 0.1 ms (Smart thing).
    & Introducing context into CP-ABE and providing dynamic access control.                               \\ \hline 
     \cite{ghosh2021case} 
    & CP-ABE   &  \textbf{RS1}\&\textbf{RS2}      &           The study mainly evaluates average network delay, average energy consumption, and average packet loss. In each metric, CASE has significant improvement by about 33\%.                      & Introducing edge intelligence to avoid manually providing
context information and sensitive data inference.                             \\ \hline 
    \cite{zheng2022adaptive} 
    & CP-ABE   &  \textbf{RS1}\&\textbf{RS2}      &           Due to the three-levels trust degree, the number attributes of is constant.                     &  Depending on AI to determine users' trust degree in different regions adaptively, and combining CP-ABE with the trust degree.                             \\ \hline 
\end{tabular}
}
\caption{General access control schemes in data storage.}\label{Table::ACStorage}
\end{table}

\begin{table*}[]
\centering
\fontsize{10pt}{11pt}\selectfont{
\begin{tabular}{|c|p{10cm}|}
\hline
Notation & Description \\ 
\hline
AAs & attribute authorities \\
\hline
n & Total number of attributes in the system \\
\hline
$\left|\mathbb{P}\right|$ & Number of attributes in access policy $\mathbb{P}$\\
\hline
$\left|\mathbb{A}\right|$ & Number of attributes in user’s attribute set $\mathbb{A}$\\
\hline
$N_{A_{p}}$ &  Number of AAs whose attributes are used in policy $\left|\mathbb{P}\right|$ \\
\hline
$N_{A_{u}}$ &  Number of AAs whose attributes are used in user's attribute set $\left|\mathbb{A}\right|$ \\
\hline
$T_{G}$ & Time to execute an exponentiation operation in an element of source groups $G_{1}$ and $G_{2}$ \\
\hline
$T_{G_{T}}$ & Time to execute an exponentiation operation in an element of target group $G_{T}$ \\
\hline
$T_{p}$ & Time to execute one pairing operation \\
\hline
$\left|Z_{p}\right|$ & Size of an element of $Z_{p}$ \\
\hline
$\left|G\right|$ & Size of an element of source group $G_{1}$ and $G_{2}$ \\
\hline
$\left|G_{T}\right|$ & Size of an element of target group $G_{T}$\\
\hline
$\mathcal{S}$, $\mathcal{R}$ & Attribute sets of a sender and a receiver \\
\hline
$\mathbb{S}$, $\mathbb{R}$ & Policies associated with a sender and a receiver \\
\hline
$N_{r}$ & Number of rules in $\left|\mathbb{P}\right|$ \\
\hline
Da &  Minimum number of attributes satisfying the access policy \\
\hline
$|G^{'}|$ & 160 bits considered as a unit of measurement \\
\hline
\end{tabular}
}
\caption{Notations.}\label{Table::Notations}
\end{table*}

The aforementioned studies provide general considerations for enhancing the performance of ABE in edge computing. Apart from these general considerations, we should have some access control schemes for specific domains. We next simply present some studies in transportation and healthcare domains. 

In the realm of transportation, 
edge computing is vital for conducting data access with low latency, and ABE-based access control is researched correspondingly. For example, OpenVDAP \cite{OpenVDAP}, an open full-stack edge computing-based platform, is dedicated to achieving low-latency data analytics for connected and autonomous vehicles (CAVs). Vehicle sensing data include historical data and real-time data, and different data sources have different access patterns. AC4AV \cite{AC4AV} first defines the data access control problem in CAVs. AC4AV can support various access control models (including ABE) by allowing users to dynamically adjust and customize access control models. AC4AV is implemented based on the OpenVDAP platform.  
To address the security and privacy issues in car-hailing systems, Sun et al. \cite{sun2022practical} propose a fine-grained puncturable matchmaking encryption (FP-ME) based on MABE and puncturable encryption. In FP-ME, not only the fine-grained access control between passengers and drivers about orders can be achieved, but also, the privacy and authenticity of passengers' orders are protected and the timeliness of passengers' ciphertext orders is guaranteed.
To achieve secure data sharing in intelligent autonomous vehicle platoons, Sun et al. \cite{sun2022secure} 
initially propose a privacy-preserving data share mechanism with flexible cross-domain authorization (FDSM-FC). In FDSM-FC, the privacy-preserving matchmaking encryption technology is combined with the cross-domain transformation primitive. In this way, it exploits the dual merits of those technologies and achieves data authenticity and authorization simultaneously.
In addition, Bao et al. \cite{9830119} present an EA-PH-ABE-DS scheme in the Internet of Vehicles (IoV) scenario where data services are expected to be more flexible and diversified. EA-PH-ABE-DS efficiently supports the update of ciphertext and user secret key in IoV and leverages edge computing to reduce computation overhead further.

In the health care domain, Zhang et al. \cite{zhang2021lightweight} design a lightweight and fine-grained access control for electronic medical record sharing systems. Therein, symmetric encryption is used to encrypt medical records by data owners and ABE is used to protect symmetric key.  
Hassan Nasiraee and Maede Ashouri-Talouki \cite{nasiraee2021privacy} propose a new privacy-preserving distributed data access control (PDAC). In PDAC, a new user anonymity approach, a novel policy-hiding mechanism, and an independent authority system are simultaneously considered for improving privacy preservation.
The anonymity approach and policy hiding can prevent untrusted authorities from colluding, and the authority system adds/removes an authority without reinitializing other authorities.

\subsection{Summary and lessons learned}
In this section, we have discussed the studies to satisfy \textbf{RS1}\&\textbf{RS2}. Some approaches are summarized in Table \ref{Table::ACStorage} for the discussion, and
learned lessons are as follows:


\begin{itemize}
    \item 
\textit{Rational use of computing power:} In these research studies, resource-constrained IoT devices leverage the computing powers from third parties (e.g., edge nodes and cloud) to offload the computing load from cryptography algorithms.
Most works simply and rigidly outsource the whole computing tasks by neighbor edge servers. 
The outsourcing model indicates that IoT devices need neighbors with relatively ample resources. 
However, the neighbors may have different computing resources. 
For example, general CPU-based edge servers can achieve outsourced computing tasks. Also, ASIC is designed for accelerating data encryption/decryption \cite{gaj2009fpga}. Moreover, the network from IoT devices to neighbors changes randomly. For example, edge servers intermittently offline, and large traffic induces network congestion \cite{EssentialsofEC}.
It is advisable to have more flexible task-offloading strategies for cryptography algorithms, which could improve the availability of computing power. Such that, resource access can be controlled more efficiently in pervasive edge computing by more rational use of computing power.

\item
\textit{Reduction in computational power requirements:} Larger key size and ciphertext size need more computing resources regardless of the usage strategies of computing power. Some research studies focus on improving the performance of cryptography-based schemes by cutting down these sizes or making them constant. 
However, these solutions reduce the expressive capability of access policy, i.e., the system's ability to satisfy ample security requirements is declined.
We should consider a trade-off between policy expressiveness and data size. 
Besides, reducing the requirement for computing power appropriately is labor-intensive and expert. 
It could facilitate the development of cryptography-based access control if we make that trade-off easier for general engineers.

\item
\textit{Introduction of additional access control components:} In different application scenarios, various components for access control are utilized, including bilateral practitioners, trust, context information, etc.
For example, edge nodes and cloud are bilateral practitioners for determining which IoT devices can access stored data. In this way, access policy is considered with more components, as a result, more useless data can be previously filtered for IoT devices by more constraints. 
The trust of data users is a factor for dynamically determining whether encrypted data can be decrypted. 
Context evaluates whether the current time or location is satisfied for decrypting data. 
In edge computing, we cannot sometimes pre-know which components are suitable for which scenario. 
Automatically recognizing components for specific scenarios can facilitate data security in pervasive edge computing.    

\item 
\textit{Reduction in human power:} 
In the studies mentioned above, some use machine learning/deep learning to substitute human power for achieving functional components in access control systems.
In this way, the components, which conventionally need human power for maintenance, can be optimized enough, and as a result, the access control system can be more efficient.
For example, Ghosh et al. \cite{ghosh2021case} largely reduce the average network delay, energy consumption, and packet loss by integrating context-aware attributes learning scheme into CP-ABE.
In addition, data privacy in the system could be improved due to less human involvement. 
Machine learning-based access control should be researched further to introduce effective and efficient intelligent algorithms into access control schemes.

\item The transportation and healthcare domains involve many sensitive data. When we design the access control schemes in the end-edge-cloud architecture in these domains, the privacy-preserving methods should be considered thoroughly.  


\end{itemize}

\begin{table*}[]
\centering
\fontsize{10pt}{11pt}\selectfont{
\begin{tabular}{|c|c|c|c|c|p{3.2cm}|c|c|}
\hline
Ref & C & O & P & M &  \multicolumn{1}{c|}{G} &Requirements & Application  \\ \hline
\cite{cao2020policy}  &  \checked  & \checked  & \checked & \checked &   Spatiality, Temporality, and Aggregation       & \textbf{RU1}\&\textbf{RU2}  & Smart cities \\ \hline
\cite{munoz2020data}  &  \checked &  \checked  &   \checked &  \checked  & Temporality and Aggregation  & \textbf{RU1}\&\textbf{RU2} & Food industry \\ \hline
\cite{kelbert2018data} & - &   \checked      &     -       &      -   &   Temporality       & \textbf{RU1}\&\textbf{RU2} & General \\ \hline
\cite{cirillo2020intentkeeper} &    -    &    -     &   \checked     &  -     & Temporality  &  \textbf{RU1}\&\textbf{RU2} & Automotive industry \\ \hline
\cite{xue2022sparkac} & \checked  &  -  &   \checked    &  -   &  Temporality and Aggregation & \textbf{RU1}\&\textbf{RU2} &  General \\ \hline
\cite{arora2022higher} & \checked  &  -  &   \checked    &  -   & - & \textbf{RU1}\&\textbf{RU2} &  General \\ \hline
\end{tabular}

 \begin{tablenotes}
     \item[1] C: Centralization, O: Obligation, P: Purpose, M: Monetization, G: Granularity
 \end{tablenotes}
}
\caption{Comparison among access control schemes in edge computing from data usage.} \label{Table::DataUsage}
\end{table*}

\section{Access Control in Data Usage}

\label{Sec::ACInDataUsage}

After the access control mechanisms in data collection and storage grant access to data, their responsibility for protecting data would end. However, some data owners could expect to learn about how their outsourced data is used and control their data usage.
Data usage control could ensure data sovereignty for data owners.
We next present the recent efforts in DUC.


Generally, DUC involves centralized and decentralized implementations. 
Cao et al. \cite{cao2020policy} implement a centralized data usage platform for smart cities, where data usage is restricted by data providers, data monetization is allowed, various obligations are considered, and the used/processed data can be visualized. Therein, the user's role can be used for authorizing data monetization (RBAC). As shown in experiments, the end-to-end delay, i.e., the time from processing the data consumer request to getting the data response, is increased by 42 ms on average. The time of processing usage control is increased by 36 ms on average.  Similarly, Munoz-Arcentales et al. \cite{munoz2020data} present a centralized implementation of DUC for securing data sharing among multiple organizations in the industry. 
Kelbert et al. \cite{kelbert2018data} formalize, implement, and evaluate a fully decentralized system of DUC, where data flow can be tracked, and data usage policies can be propagated and enforced effectively. The assessment indicates that 1) dataflow tracking and policy propagation achieve 21\%–54\% of native execution throughput, and 2) decentralized policy enforcement generally outperforms centralized methods.

Traditionally, the DUC mechanism pre-defines data usage pipelines across domains, e.g., IoT-generated data is outsourced to edge servers for advanced analysis. However, this fashion complicates the process of DUC policy specification, especially when the data usage pipeline is mutable. The existing DUC  techniques cannot preventively and proactively enforce data usage control.
Cirillo et al.
\cite{cirillo2020intentkeeper} propose a practical distributed DUC framework IntentKeeper based on the open-source IoT edge computing framework FogFlow \cite{cheng2017fogflow}. It decouples data usage flow with data usage policy and allows users to specify DUC policy easily using an ODRL-like model \cite{ODRL}. And, it supports preventive and proactive enforcement by federated service orchestration and access control data management. Use case validations demonstrate that IntentKeeper reduces the complexity of policy specification by up to 75\% in moderately complex scenarios, compared with current flow-based approaches. Additionally, experimental results indicate that IntentKeeper ensures rapid response times (under 40 ms) with minimal overhead (below 10 ms).



To enforce access control on the general big data computing platform, Spark, Xue et al. \cite{xue2022sparkac} propose purpose-aware access control by enhancing purpose-based access control with data operation/processing purpose (describing the fine-grained data usage). They implement purpose-aware access control in the execution and optimization layers. As a result, the SQL query, machine learning algorithm, streaming computation, and graph computation can be automatically supported by this solution, meanwhile introducing acceptable latency for big data analytics. The fine-grained purpose and the implementation methodology can be adapted for other data analytics at the edge. 
In addition, IntentKeeper \cite{cirillo2020intentkeeper} supports vehicle analytics-related purposes for enforcing DUC.

Arora et al. \cite{arora2022higher} use the higher-order relationship context information to control data usage in the dynamic environment in IoT applications. This scheme uses the history of relationship changes to make authorization decisions. This means that the relationship temporality should be considered. For example, if a nurse has already used the medical data of Alice for care, this nurse cannot use the medical data of Bob for medical research purposes. The proposed graph-matching algorithms for authorization are effective and efficient as shown in the experiments when temporality is introduced into ReBAC.


\textit{Summary and Lessons Learned:}
In this section, we have discussed the studies to satisfy
\textbf{RU1}\&\textbf{RU2}. The approaches are summarized in Table \ref{Table::DataUsage}, and the learned lessons are as follows:
Most schemes focus on a specific application realm to implement centralized DUC by building the corresponding purpose set. 
Obligation is considered in most DUC implementations.
DUC can use monetization to facilitate data flow. 
The granularity of DUC can be divided into three categories: spatiality, temporality, and aggregation. 
Different schemes may involve different granularity.
The purpose component is regarded as the core component for satisfying data owners' privacy in DUC \cite{akaichi2022usage}.  
In purpose-based schemes \cite{byun2008purpose}, data owners specify ``which data users can use data in which purpose''. If the usage purpose indicated by data users complies with the data-owner-specified purpose, the user's query will be allowed to use data. 
Purpose-based schemes control data usage before execution, meaning that the data analysis logic operates on data loaded into memory in processors (e.g., CPU).

\section{Blockchain-based Access Control Platform}
\label{Sec::BBACP}
Although the studies in the above sections can be selected for achieving access control, which depends on small resources or has acceptable delay, flexibility, and scalability, they do not consider the security of critical access control data in edge computing.
In this section, we study how to ensure this security by depending on the advantages of blockchain, i.e., decentralization, immutability, programmability, and security  \cite{dabbagh2021survey, zhang2021resource, zheng2018blockchain}.
We introduce the recent blockchain-based access control platforms developed for managing the data of access control in different data lifecycles. In particular, we mainly introduce which typical access control models are combined with blockchain, and the method of exploiting blockchain to manage the data of these models.

\subsection{Blockchain-based access control in data collection}

Liu et al. \cite{liu2020fabric} introduce an access control system, Fabric-IoT, which exploits Hyperledger Fabric to manage ABAC used for the resource data generated by IoT devices. They use smart contracts to implement resource access, policy management, and policy enforcement. In particular, the device contract provides methods to store/query URLs of resource data; the policy contract provides functions to manage access control policies; and the access contract enforces access control. 
The simulation experiments present acceptable throughput and efficient consensus.

Mazzocca et al. \cite{mazzocca2022framh} propose a risk-aware authorization middleware for healthcare based on federated learning. 
Blockchain stores the trained federated model, which evaluates the risk level of the current patient. 
Sylla et al. \cite{sylla2021blockchain} propose a new blockchain-enabled decentralized context-aware authorization system. 
Users can dynamically define contextual access rights for each IoT application in smart contracts. 
When an authorized client issues a request, the smart contract generates the corresponding contextual access token. With the edge deployment of blockchain, the maximum generation time of contextual access tokens is about 10 ms. 

In practice, we need the management of capability registration, delegation, and revocation in CapBAC.
The centralized management component likely induces performance bottlenecks and single-failure problems. 
The blockchain can be used for capability registration, delegation, and revocation \cite{liu2021capability, xu2019exploration}.
The blockchain-enabled decentralized CapBAC (BlendAC), which is being researched in the academic community, can avoid those problems. 
Xu et al. \cite{xu2018blendcac} propose a BlendAC for large-scale IoT systems. It achieves robust identity-based capability token management, where the smart contract is used for registration, validation, delegation, and revocation. They implement BlendAC on a Rasberry Pi device and on a local private blockchain. The experimental results demonstrate that BlendAC can offer decentralized, scalable, lightweight, and fine-grained access control.
Similarly, the following studies design the blockchain platforms based on CapBAC, where different capability tokens are considered.
IoT consortium capability-based access control (IoT-CCAC) \cite{bouras2021iot} utilizes blockchain to achieve decentralized CapBAC schemes for consortium applications in IoT cities, where IoT-generated data from different parties is governed for different services by using group capability tokens.
Nakamura et al. \cite{nakamura2020exploiting} propose a novel CapABC scheme, where the smart contract is used for managing the tokens that each combines with an action and are different from the conventional token model, i.e., one token per subject. This solution allows subjects to have multiple tokens and facilitates flexible token delegation among multiple subjects. Chen et al. \cite{chen2023capability} present a fine-grained and flexible access control scheme, which simultaneously achieves flexible capability granting and revocation, sufficient capability granularity, secure capability delegation, and capability verification. This scheme combines blockchain with CapBAC, where a smart contract manages subject and object information.
It designs authorization rules and a tree for capability, where the token considers context information.


\subsection{Blockchain-based access control in data storage}
Considering the difficulty of deploying the existing blockchain platforms, Zhang et al. \cite{zhang2021lightweight1} design a lightweight blockchain (LBC) by improving the proof-of-work consensus for lightly implementing blockchain. Moreover, they use blockchain to store access control policy to protect the privacy of policy and design six smart contracts for integrating ABAC and ABE. 
In particular, the policy management contract (SCPA) operates the policy set and generates encryption policies; the policy decision contract determines the policy compliance according to policies from SCPA; the policy information contract manages attribute information for SCPA; the encryption and decryption contract enforces ABE by edge nodes; the penalty mechanism contract manages the punitive measures for illegal access; the policy enforcement contract returns the encrypted data or available resources.
Attribute updates/revocations in ABE are incompatible with the immutability of blockchains.
To address this issue, Yu et al. \cite{yu2020enabling} propose a new blockchain-based access control for IoT systems, which allows attribute updates/revocations by integrating Chamelon Hash (CH) algorithms into blockchains. To ensure the security of key updates, they further design a key chain for verifying key updates to avoid the abuse of CH update function.
However, these works do not consider the privacy preservation of user data. Han et al. \cite{han2021blockchain} propose a blockchain-based auditable access control system for private data in IoT. Based on the ABAC model, they define a model to divide the IoT data into private data and public data and utilize smart contracts for storing and managing the private data, access control policy, and access records. 
In the smart grid area,
Yang et al. \cite{yang2020secure} present a blockchain-based CP-ABE access control system for smart grid, where amounts of data generated by IoT devices need to be shared among distributed energy resources, microgrids, and the main grid. The blockchain is combined with edge computing, where the partial encryption/decryption tasks of CP-ABE are executed in the form of smart contracts. 


\subsection{Blockchain-based access control in data usage}
IntentKeeper \cite{cirillo2020intentkeeper} uses Hyperledger Fabric to propagate, synchronize, and record some important data, e.g., data usage policies, distributed sites, and policy enforcement decisions.
Syed et al. \cite{syed2020novel} design a blockchain-based vehicle lifecycle tracking framework, where Hyperledger Fabric enables the DUC engine to monitor vehicle maintenance, insurance, and rented vehicles continuously (e.g., restriction on the number of vehicle leases). Specifically, they develop smart contracts for enforcing DUC and maintaining DUC policy.
Wu et al. \cite{9662435} present the TS-PBAC model, which leverages blockchain-enabled secure logs to evaluate quality data from the external network. This model allows individual-centric privacy preservation in the Internet of Medical Things scenarios, where medical data can be specified for different purposes by data owners. Meanwhile, local differential privacy and RBAC are combined to protect sensitive attributes against internal users. 
To mitigate the trust and security problem for IoT big data management, Ma et al. \cite{zhaofeng2019blockchain} propose a BlockBDM scheme, where public and permissioned blockchain is utilized for performing data operations and their management. In particular, BlockBDM achieves secure usage control by limiting the operations (e.g., data backup and data save) and recording data usage for auditing trails. 
These studies \cite{xiao2020privacyguard, zhang2021data, pol2021preserving,gao2021blockchain,han2022blockchain} combine trusted execution environment (TEE), e.g., SGX, with blockchain. The smart contract is used to manage data usage policies and data trading prices. Controlling data usage is achieved by a TEE-based off-chain contract execution engine, and the results are securely committed to the blockchain for the auditing trail.

\subsection{Blockchain-based versatile access control platform}
A blockchain-based versatile access control platform can provide multiple access control implementations, which intends to be used for multiple data lifecycles.
Implementing a useful versatile access control platform is tricky because data and services are complicated in edge computing, the application scenarios cannot be pre-known, and on-chain data is costly \cite{international2018new}. 
Focusing on the smart contract-based access control (SCAC), Zhu et al. \cite{zhu2022fine} explore how to combine blockchain security with fine-grained access control models while keeping the efficiency of edge computing. SCAC depends on ABAC and RBAC to develop basic entities in the model of SCAC, including attributes (e.g., token and identity), smart contracts (e.g., register manager and device, set attribute predicate, and query information). 
Ouaddah et al. \cite{ouaddah2016fairaccess} present a blockchain-based access control framework FairAccess to enable its users to own and control their data. It uses a public Bitcoin-like blockchain to act as the access control manager, which supports various access control models (e.g., RBAC and ABAC) and transactions (grant, get, delegate, and revoke access). In \cite{ouaddah2021fairaccess2}, they update FairAccess to FairAccess2, where more granular data access control policies are supported based on Ethereum. Based on Ethereum, Oscar et al. \cite{novo2018blockchain} propose a decentralized access control system for the IoT. This system utilizes blockchain technology to store and distribute access control information, and it supports resource-constrained devices. A single smart contract is designed to enforce access control based on the access control policy of a set of IoT devices, where a manager is supervised to manage these devices. However, this access control approach is too complex for changing permissions and cannot fully meet the dynamic requirements of the IoT.
Zhang et al. \cite{zhang2018smart}  present a novel framework for the IoT that centers around Access Control Contracts (ACCs). In this framework, each ACC is responsible for defining a specific access control method based on a static access control policy. Simultaneously, a Judge Contract dynamically evaluates the ACCs by assessing the behavior of subjects registered in the system. The Register Contract records access control information, misbehavior-judging methods, and the associated smart contracts, offering functionalities such as registration and updates. This approach provides a structured and dynamic means of managing access control within IoT environments.

\subsection{Summary and lessons learned}
In this section, we have discussed the studies applying blockchain to protect important access control data while satisfying \textbf{RC1}\&\textbf{RC2}, \textbf{RS1}\&\textbf{RS2}, or \textbf{RU1}\&\textbf{RU2}. 
The approaches are summarized in Table \ref{Table::BlockchainForAC}, and
the learned lessons are as follows:

\begin{table*}[]
\centering
\fontsize{10pt}{11pt}\selectfont{
\begin{tabular}{|p{0.5cm}<{\centering}|p{0.6cm}<{\centering}|p{1.85cm}<{\centering}|p{1.7cm}<{\centering}|p{7.2cm}|}
\hline
LC                                & Ref               & Blockchain         & Supported model(s) & Operations                                                                                             \\ \hline
\multirow{5}{*}{DC}      & \cite{liu2020fabric}             & Hyperledger Fabric & ABAC               & Policies CRUD, policies enforcement, resource URLs CRUD                                                \\ \cline{2-5} 
                                      & \cite{mazzocca2022framh}             & Ethereum & RAAC               &  Stores the trained federated model
                                              \\ \cline{2-5} 
                                      & \cite{sylla2021blockchain}             & Hyperledger Fabric, Ethereum & CAAC, CapBAC              & Context token register, validation and revocation                                               \\ \cline{2-5} 
        &\cite{xu2018blendcac}             & Ethereum           & CapBAC             & Identity token register,validation, delegation, and revocation                                         \\ \cline{2-5} 
                                      & \cite{bouras2021iot}             & BigchainDB         & CapBAC, GBAC             & Group token register,validation                                                                        \\ \cline{2-5} 
                                      & \cite{nakamura2020exploiting}            & Ethereum           & CapBAC             & Action token register, validation, delegation, and revocation                                           \\ \cline{2-5} 
                                      & \cite{chen2023capability}            & Hyperledger Fabric & CapBAC, CAAC             & Token register, validation, delegation, and revocation                                          \\ \hline
\multirow{4}{*}{DS} & \cite{zhang2021lightweight1}  & LBC                & CP-ABE, ABAC      & ABAC policy CRUD, and encrypting/decrypting data or policy file by CP-ABE                              \\ \cline{2-5} 
                                      & \cite{yu2020enabling}                   & -                & CP-ABE             & Attributes update/revocation                                                                                     \\ \cline{2-5} 
                                      &  \cite{han2021blockchain}                   & Hyperledger Fabric & ABAC               & Upload/Query private data and URLs of public data, Update/Query access record, Policy CRUD             \\ \cline{2-5} 
                                      & \cite{yang2020secure}   & Hyperledger Fabric & CP-ABE             & Data owners and data users can use smart contracts to accomplish partial data encryption and decryption \\ \hline
\multirow{2}{*}{DU}        & \cite{cirillo2020intentkeeper}  & Hyperledger Fabric & DUC               & Upload/Query data usage policies, distributed sites, policy enforcement decisions.                     \\ \cline{2-5}
                                      &  \cite{syed2020novel}        & Hyperledger Fabric & DUC                & Upload/Query data usage policies and attributes, DUC enforcement engine                                \\ \cline{2-5}
                                       &  \cite{9662435}        & ChainSQL & DUC                & Record EMR records, model parameters of access control, privacy policies, and access logs. Enforce PBAC                                \\ \cline{2-5}
                                       & \cite{zhaofeng2019blockchain}          & Hyperledger Fabric  & DUC                & Limit the number of uses, the period of use, backups, etc.                                \\ \hline
\multirow{2}{*}{ML}        & \cite{zhu2022fine}  & Hyperledger Fabric  & RBAC, ABAC               & Register manager and device, set attribute predicate, query information, etc.                     \\ \cline{2-5}
                                      &  \cite{ouaddah2016fairaccess, ouaddah2021fairaccess2}        & Modified Bitcoin, Ethereum&  RBAC, ABAC, etc.               & Grant, get, delegate and revoke access                                \\ \cline{2-5}
                                       &  \cite{novo2018blockchain}        & Ethereum & General 
                                       &  Register manager and device, add and remove manager, add access control enforcement, revoke permission, etc.                             \\ \cline{2-5}
                                       & \cite{zhang2018smart}          & Ethereum  & General 
                                       & Register
access control information, misbehavior-judging method and
corresponding smart contract, and provides operations (e.g.,
register and update)                               \\ \hline
\end{tabular}

  \begin{tablenotes}
     \item[1] LC: Lifecycle, ML: Multiple Lifecycles
   \end{tablenotes}
}
\caption{Comparison among blockchain-based access control schemes in edge computing for data collection, storage, and usage.} \label{Table::BlockchainForAC}
\vspace{-0.1cm}
\end{table*}

\begin{table*}[]
\centering
\fontsize{10pt}{11pt}\selectfont{
\begin{tabular}{|p{0.8cm}<{\centering}|p{2cm}<{\centering}|p{3.8cm}<{\centering}|p{5cm}|}
\hline
 Ref               & Blockchain         & Supported model(s) & Operations                                                                                             \\ \hline
\cite{zhu2022fine}  & Hyperledger Fabric  & RBAC and ABAC               & Register manager and device, set attribute predicate, query information, etc.                     \\ \hline
\cite{ouaddah2016fairaccess, ouaddah2021fairaccess2}        & Modified Bitcoin, Ethereum&  RBAC, ABAC, etc.               & Grant, get, delegate, and revoke access                                \\ \hline
\cite{novo2018blockchain}        & Ethereum & General architecture for resource-constrained devices  &  Register manager and device, add and remove manager, add access control enforcement, revoke permission, etc.                             \\ \hline
\cite{zhang2018smart}          & Ethereum  & General architecture for resources access in IoT                & Register
access control information, misbehavior-judging method and
corresponding smart contract, and provides operations (e.g.,
register and update)                               \\ \hline
\end{tabular}
}
\caption{Comparison among versatile blockchain-based access control schemes in edge computing.} \label{Table::BlockchainForAC1}
\vspace{-0.1cm}
\end{table*}

\begin{itemize}
    \item 
In data collection, many studies focus on utilizing blockchain technology to optimize traditional CapBAC by implementing decentralized CapBAC. The smart contract generally accomplishes operations on tokens in CapBAC. 
As lessons learned in Section \ref{Sec::SALInDataCollection}, the tokens in CapBAC are often incorporated with some components (e.g., trust and group).
As such, it would be interesting to build a blockchain management system, which can provide a library of various tokens for complex edge computing applications.

\item 
In data storage, blockchain can be used to manage access policies in ABE, execute computing in ABE, record access permissions, etc. Depending on a blockchain, we can integrate ABAC and ABE securely while achieving flexible and scalable access control.

\item 
Blockchain in each DUC scheme is generally used as a platform to regulate data usage rules, guarantee the consensus of executing data usage rules, and record model parameters and data usage records.
Purpose-based schemes generally request data users to input their data usage purpose. 
In this fashion, malicious users may cheat purpose-based enforcers. 
Relational audits can deter data users from being dishonest if the purpose-based schemes are integrated into blockchain to manage purpose.

\item 
From the blockchain for access control in data collection, storage, and usage, we can observe that each data phase has a limited number of blockchain-supported models.  
Given the intricate and dynamic nature of the access control requirements, each data phase should be equipped with a sufficient range of blockchain-supported models.
In addition, we observe that the access control schemes in each data phase have common components, such as model parameters management, policy CRUD, access logs record, attributes CRUD, model enforcement, etc.
We could design a lightweight blockchain-based versatile access control platform depending on these common components.
Finally, these blockchain-based approaches are crafted for distinct scenarios individually, involving Hyperledger Fabric, Ethereum, ChainSQL, etc.  
If each separated blockchain in each data phase integrates appropriately, it could induce extra overhead, such as communications between different access control systems of multiple data phases, and prevent access control performance from being improved in edge computing.
The interaction among different blockchains should be supported for access control systems.

\item 
Blockchain can support multiple access control models for multiple lifecycles, where smart contracts enable various operations: register access control enforcement, manage attributes, etc. 
Versatile access control technology could be the feasible solution to solve the data security problem in edge computing, which has various stakeholders, ample data sources, diversified dynamics, and complex applications. However, the current blockchain-based versatile access control platforms mainly support conventional models (e.g., ABAC, RBAC, and CapBAC), and lag the development of new techniques as introduced in Section \ref{Sec::ACInDataCollection}\&\ref{Sec: ACInDataStorage}\&\ref{Sec::ACInDataUsage}, where the improvements for conventional models are designed.
\end{itemize}

\section{Challenges and Future Research Directions}
\label{Sec::Discussions}

Apart from the aforementioned issues, there remain challenges and new research directions in edge computing access control to be discussed, as follows.

\subsection{Machine learning-based access control in IoT edge computing}

Traditional access controls commonly depend on predefined rules and require laborious maintenance and updates. With the widely distributed resources in edge computing, the demand for a more dynamic and efficient access control system is escalating. 
Machine learning can play an important role in satisfying the demand \cite{nobi2022towards}, where trained machine learning models constitute some blocks in access control systems to automate and reduce labor. Some studies use machine learning technology in their access control systems, e.g., \cite{atlam2021fuzzy, zheng2022adaptive, ghosh2021case}. 
We would discuss and envisage future research directions in the following aspects.

\begin{itemize}
    \item 
\textbf{Limited datasets:} 
A complete access control policy is the prerequisite for effective data protection.
The natural language documents for access control requirements are essential sources of access control policies \cite{narouei2018automatic}. Natural language processing technology is the common tool for extracting access control policies from those documents. It identifies and analyzes policy information described in human language, and converts this information into structured data that computers can comprehend and process.  
For example, Freisleben et al.\cite{ait2017abac} present an approach based on K-nearest neighbors algorithms for clustering ABAC policies, which can reduce the number of policy rules by evaluating similarity. Text2Policy \cite{xiao2012automated} classifies sentences into four access control modes. If a sentence matches, it extracts the subject, action, and object from the sentence by using the comments of the sentence. Text2Policy does not support contextual information extraction, nor does it support complex sentence patterns. To overcome the above defects, Narouei et al. \cite{narouei2018automatic} improve semantic role labeling and introduce domain-specific knowledge to greatly improve the performance of extracting implementable policies from natural language access control policy documents. The above work can support RBAC, ABAC, and Access Control Lists, but they do not support DUC. Hosseinzadeh et al. \cite{hosseinzadeh2020systematic} systematically propose a method to extract implementable policies from data usage requirement documents from the perspectives of data owners: policy declaration, policy transformation, and policy negotiation.

\textit{Limitation:} These studies facilitate the development of techniques for extracting access control policies from natural language documents. However, the explored datasets are mostly from conventional domains such as education and medicine. The lack of datasets from edge computing, especially in IoT, could impede the development of its access control technologies.
Besides, real-world access control policies generally involve sensitive data, which obstacles the development of datasets.  

\textit{Possible Solutions:} Policy generation can mitigate the limitation \cite{nobi2022machine}. The existing approaches to generating policy are problematic, e.g., random policy generation and manual policy writing. The datasets created using these methods are idealized, and do not fully capture the complex authorization scenarios in the real world. To generate simulated data with a similar distribution to any realistic data, 
we can explore pre-trained models (e.g., GPT4 \cite{achiam2023gpt}), such that the study of machine-based access control can reduce the reliance on real-world open-source datasets.
The following aspects should be considered: 1) access control models, 2) application domains, and 3) different neural networks for data simulation.
Different access control models have different organizations about access control information so the pre-trained models have different performances on different models. 
Many applications in IoT, e.g., the Internet of Vehicles and smart homes, have different data flow and business parties. Different application domains require different data simulations.
In large-scale access control scenarios, business parties can leverage pre-trained models to simulate data, test the access control functions overall beforehand, and monitor access control systems timely. Different neural networks require different computation power and could have different performances for various access control models and application scenarios. To make a cost-effective data simulation, business parties should select appropriate pre-trained models. 
\item

\textbf{Federated learning:} Different business organizations in the edge computing ecosystem would form alliances based on business interests. The aforementioned studies in data lifecycles can be used intra-alliance, in which data owners and consumers cooperate to exert IoT-generated data. Due to data owners' limited data protection ability, their access control policies may be incomplete. These data owners can use the access control policies of other data owners to enhance their access control policies, such as adding missed access control policies or updating inferior access control policies. For example, Abu Jabal et al. \cite{abu2023flap} propose FLAP, a federal learning framework for ABAC. It allows data owners to learn ABAC policies from each other, ensuring that each party generates accurate policies. Moreover, FLAP takes into account policy conflicts and privacy concerns. To solve the problem of insufficient access control policies in a single smart home IoT context, Yu et al. \cite{yu2020learning} propose the federal multi-task learning framework LoFTI for several federated smart homes. The framework ensures the privacy of the access control policy in a smart home, enhances its context-based access control policy, and realizes the personalization of the smart home policy. Federated learning could be feasible in the cross-domain optimization of access control policies. However, they have limited consideration of the complex data lifecycles in edge computing. Multi-data lifecycles indicate access control policies of multiple access control models. An interesting research direction is to optimize each data owner's access control policies of multiple access control models, meanwhile without privacy leaking among data owners. A possible solution could be versatile federated learning, targeting a multi-access control federated model, which allows data owners to enhance access control policy with less consideration of the access control models.

\item
\textbf{Adversarial attacks:} Although machine learning-based access control systems have the potential for enhancing data security in largely distributed edge computing, the deployment of such systems can be hampered by the vulnerability of machine learning models to adversarial attacks. These attacks exploit weaknesses in machine learning algorithms to violate the system's access control policies to gain unauthorized access. 
For example, Nobi et al. \cite{nobi2022adversarial} propose that adversarial attacks can be reduced to a certain extent by utilizing specific constraints on access control. We would further explore solutions for implementing machine learning-based access control and mitigating various attacks within the edge computing ecosystem. These solutions should consider the characteristics and features of access control in edge computing, such as various models and large scale. 

\end{itemize}

\subsection{Hybrid access control}

\begin{figure}
	\begin{center}
	\includegraphics[scale = 1.2]{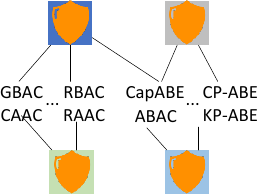}
		\caption{Hybrid access control.}
		\label{Fig::HybridAccessControl}
	\end{center}
 \vspace{-0.2cm}
\end{figure}

Hybrid access control technology can improve access control performance by combining the advantages of different access control mechanisms \cite{pal2021protocol}. In edge computing access control, the introduced models in Section \ref{Sec::TraAccessControl}, e.g., CapABC, ABE, and DUC, are a set of data security techniques.  As shown in Figure \ref{Fig::HybridAccessControl}, we can combine these models' components to enhance the performance of access control in edge computing.
For example, to satisfy the low-latency requirements and secure access control systems for the Industrial Internet-of-Things (IIoTs). DHACS \cite{saha2021dhacs} combines these elements: role, context, and organization regulation. The functions of access control are encoded in smart contracts. The operational transactions are recorded on the blockchain. The efficiency of access control management is improved by about 30\%. DHACS is the first attempt to use a decentralized blockchain with smart contracts for hybrid access control in IIoTs. 
Because services involve data analytics, it is reasonable to introduce appropriate components of DUC into CapBAC to optimize access control performance further. The components of purpose-based schemes can be applied to ABE-based access control, which could further restrict data usage in advance and reduce the computing cost of decryption by purpose.

Hybrid schemes' effectiveness, efficiency, and flexibility can be researched for high-performance edge computing access control under pervasive data analytics.
For example, the method to make a more flexible CapBAC should be inspected when CapBAC is combined with DUC. The method to implement efficient decryption/encryption could be investigated when the purpose attribute is integrated into the ABE scheme. 

Finally, while hybrid access control can improve performance, privacy issues in hybrid models should be investigated because hybrid means that some cases involve organization-across attributes.

\subsection{Powerful access control technology testbed}

Some testbeds in the mentioned studies use simulation tools to verify their ideas. Some deploy complex real end-edge-cloud platforms, which enable the tests to be closer to real production scenarios. 
The new paradigm, i.e., edge computing \cite{EssentialsofEC}, has complex network topology (e.g., LAN, MAN, and WAN) and heterogeneous network devices (e.g., Cisco or Huawei switch). Meanwhile, IoT devices and edge nodes are heterogeneous with different computing, storage, and communication capabilities. Except for the sophisticated hardware conditions, heterogeneous IoT and network protocols coexist. For example, MQTT and XMPP are usually used to collect IoT-derived data in a certain security domain. WiFi and 5G enable devices to connect with and communicate in networks. 
Because this physical ecosystem has enormous economic expenditures and complex maintenance, it is difficult to achieve flexible and extendable tests for research and real production scenarios.

To tackle these challenges, many open-source simulation tools for edge computing are developing \cite{mahmud2022ifogsim2, nandan2019iotsim, sonmez2018edgecloudsim}, which prompts the development of edge computing technology.  
They can simulate common infrastructures, including networks, mobile devices, edge data centers, microservices, edge orchestrators, edge clusters, and central cloud. Moreover, they support various computing loads, different data distributions, and access to real data.

These simulation systems do not consider the simulation of access control currently natively. However, they support powerful module customization and extension. These existing simulation systems can be extended to support access control simulation.
To achieve this simulation in an edge computing ecosystem, similar to supporting various network protocols, these simulation systems should flexibly integrate heterogeneous access control models/protocols. For example, the attribute authorities and context information should be allowed according to the user's configuration in the simulation system.
It is not advisable to embed rigid access control models into simulation systems according to real experiences about applications, because the applications in edge computing are various and could be changed frequently. To integrate different access control models flexibly, we need a universal library for various model modules, each of which acts as a basic building block for constructing a functional model template. 
In addition, real data should be flexibly and advisably injected for the corresponding function verification in the light of real-world situations. 
Because access control would run on enormous data in real situations, capturing all real data in a simulation system is impractical.
The trade-off between the amount of ingested data and verification accuracy should be researched, especially considering diversified data categories in edge computing. 
Finally, the high usability of the simulation system would be an important problem to be solved, so that the system users can build the target access control system efficiently. As such, researchers or technicians can focus on verifying their ideas for implementing access control functions.

\section{Conclusion}
\label{Sec::SumAndCon}

This paper has examined access control models applicable to IoT edge computing to facilitate resource-conserving, low-latency, flexible, and scalable access control. By systematically classifying these models and evaluating their effectiveness across multiple data lifecycles, we have provided some novel insights, particularly highlighted in each summary section and future research directions. This can complement existing surveys and promote readers' understanding of access control technologies in the context of IoT edge computing.
Through thoroughly exploring existing access control models and their adaptations to IoT edge computing, the survey not only presents its readers with the state-of-the-art but also emphasizes the ongoing necessity for innovation. 
Given the rapidly evolving landscape of IoT and edge computing,  it is imperative to create adaptive, efficient, and secure access control solutions. Continued research in this area is crucial to fully harness the benefits of edge computing while ensuring robust data security.
\bibliographystyle{elsarticle-num}
\bibliography{elsarticle-template-num}

\begin{thebibliography}{100}
\providecommand{\url}[1]{#1}
\csname url@samestyle\endcsname
\providecommand{\newblock}{\relax}
\providecommand{\bibinfo}[2]{#2}
\providecommand{\BIBentrySTDinterwordspacing}{\spaceskip=0pt\relax}
\providecommand{\BIBentryALTinterwordstretchfactor}{4}
\providecommand{\BIBentryALTinterwordspacing}{\spaceskip=\fontdimen2\font plus
\BIBentryALTinterwordstretchfactor\fontdimen3\font minus
  \fontdimen4\font\relax}
\providecommand{\BIBforeignlanguage}[2]{{%
\expandafter\ifx\csname l@#1\endcsname\relax
\typeout{** WARNING: IEEEtran.bst: No hyphenation pattern has been}%
\typeout{** loaded for the language `#1'. Using the pattern for}%
\typeout{** the default language instead.}%
\else
\language=\csname l@#1\endcsname
\fi
#2}}
\providecommand{\BIBdecl}{\relax}
\BIBdecl

\bibitem{gomez2019internet}
C.~Gomez, S.~Chessa, A.~Fleury, G.~Roussos, and D.~Preuveneers, ``{Internet of
  Things for enabling smart environments: A technology-centric perspective},''
  \emph{Journal of Ambient Intelligence and Smart Environments}, vol.~11,
  no.~1, pp. 23--43, 2019.

\bibitem{chen2019deep}
J.~Chen and X.~Ran, ``Deep learning with edge computing: A review,''
  \emph{Proceedings of the IEEE}, vol. 107, no.~8, pp. 1655--1674, 2019.

\bibitem{shi2016edge}
W.~Shi, J.~Cao, Q.~Zhang, Y.~Li, and L.~Xu, ``{Edge computing: Vision and
  challenges},'' \emph{IEEE Internet of Things Journal}, vol.~3, no.~5, pp.
  637--646, 2016.

\bibitem{ferrari2010access}
E.~Ferrari, \emph{Access control in data management systems}.\hskip 1em plus
  0.5em minus 0.4em\relax Morgan \& Claypool Publishers, 2010.

\bibitem{custers2019eu}
B.~Custers, A.~M. Sears, F.~Dechesne, I.~Georgieva, T.~Tani, and S.~Van~der
  Hof, \emph{EU personal data protection in policy and practice}.\hskip 1em
  plus 0.5em minus 0.4em\relax Springer, 2019, vol.~29.

\bibitem{hernandez2013distributed}
J.~L. Hern{\'a}ndez-Ramos, A.~J. Jara, L.~Marin, and A.~F. Skarmeta,
  ``{Distributed capability-based access control for the Internet of Things},''
  \emph{Journal of Internet Services and Information Security}, vol.~3, no.
  3/4, pp. 1--16, 2013.

\bibitem{ferraiolo2001proposed}
D.~F. Ferraiolo, R.~Sandhu, S.~Gavrila, D.~R. Kuhn, and R.~Chandramouli,
  ``{Proposed NIST standard for role-based access control},'' \emph{ACM
  Transactions on Information and System Security}, vol.~4, no.~3, pp.
  224--274, 2001.

\bibitem{hu2013guide}
V.~C. Hu, D.~Ferraiolo, R.~Kuhn, A.~R. Friedman, A.~J. Lang, M.~M. Cogdell,
  A.~Schnitzer, K.~Sandlin, R.~Miller, K.~Scarfone \emph{et~al.}, ``{Guide to
  attribute based access control (ABAC) definition and considerations
  (draft)},'' \emph{NIST special publication}, vol. 800, no. 162, pp. 1--54,
  2013.

\bibitem{sandhu2003usage}
R.~Sandhu and J.~Park, ``Usage control: A vision for next generation access
  control,'' in \emph{Computer Network Security: Second International Workshop
  on Mathematical Methods, Models, and Architectures for Computer Network
  Security, MMM-ACNS 2003, St. Petersburg, Russia, September 21-23, 2003.
  Proceedings 2}.\hskip 1em plus 0.5em minus 0.4em\relax Springer, 2003, pp.
  17--31.

\bibitem{park2004uconabc}
J.~Park and R.~Sandhu, ``{The UCONABC usage control model},'' \emph{ACM
  Transactions on Information and System Security}, vol.~7, no.~1, pp.
  128--174, 2004.

\bibitem{GBAC}
``{Secure access control using groups on Azure AD},''
  \url{https://learn.microsoft.com/en-us/azure/active-directory/develop/secure-group-access-control}.

\bibitem{toninelli2006semantic}
A.~Toninelli, R.~Montanari, L.~Kagal, and O.~Lassila, ``A semantic
  context-aware access control framework for secure collaborations in pervasive
  computing environments,'' in \emph{International Semantic Web
  Conference}.\hskip 1em plus 0.5em minus 0.4em\relax Springer, 2006, pp.
  473--486.

\bibitem{atlam2020risk}
H.~F. Atlam, M.~A. Azad, M.~O. Alassafi, A.~A. Alshdadi, and A.~Alenezi,
  ``Risk-based access control model: A systematic literature review,''
  \emph{Future Internet}, vol.~12, no.~6, p. 103, 2020.

\bibitem{fong2011relationship}
P.~W. Fong, ``Relationship-based access control: protection model and policy
  language,'' in \emph{Proceedings of the first ACM Conference on Data and
  Application Security and Privacy}, 2011, pp. 191--202.

\bibitem{almenarez2005trustac}
F.~Almen{\'a}rez, A.~Mar{\'\i}n, C.~Campo, and C.~Garc{\'\i}a~R, ``{TrustAC:
  Trust-based access control for pervasive devices},'' in \emph{Security in
  Pervasive Computing: Second International Conference, SPC 2005, Boppard,
  Germany, April 6-8, 2005. Proceedings 2}.\hskip 1em plus 0.5em minus
  0.4em\relax Springer, 2005, pp. 225--238.

\bibitem{mahalle2013fuzzy}
P.~N. Mahalle, P.~A. Thakre, N.~R. Prasad, and R.~Prasad, ``{A fuzzy approach
  to trust based access control in Internet of Things},'' in \emph{Wireless
  VITAE 2013}.\hskip 1em plus 0.5em minus 0.4em\relax IEEE, 2013, pp. 1--5.

\bibitem{kayes2020survey}
A.~Kayes, R.~Kalaria, I.~H. Sarker, M.~S. Islam, P.~A. Watters, A.~Ng,
  M.~Hammoudeh, S.~Badsha, and I.~Kumara, ``A survey of context-aware access
  control mechanisms for cloud and fog networks: Taxonomy and open research
  issues,'' \emph{Sensors}, vol.~20, no.~9, p. 2464, 2020.

\bibitem{zhang2018survey}
P.~Zhang, J.~K. Liu, F.~R. Yu, M.~Sookhak, M.~H. Au, and X.~Luo, ``A survey on
  access control in fog computing,'' \emph{IEEE Communications Magazine},
  vol.~56, no.~2, pp. 144--149, 2018.

\bibitem{alnefaie2021survey}
S.~Alnefaie, S.~Alshehri, and A.~Cherif, ``{A survey on access control in IoT:
  models, architectures and research opportunities},'' \emph{International
  Journal of Security and Networks}, vol.~16, no.~1, pp. 60--76, 2021.

\bibitem{ABESurvey}
M.~Rasori, M.~L. Manna, P.~Perazzo, and G.~Dini, ``{A survey on attribute-based
  encryption schemes suitable for the Internet of Things},'' \emph{IEEE
  Internet of Things Journal}, vol.~9, no.~11, pp. 8269--8290, 2022.

\bibitem{akaichi2022usage}
I.~Akaichi and S.~Kirrane, ``Usage control specification, enforcement, and
  robustness: A survey,'' \emph{arXiv preprint arXiv:2203.04800}, 2022.

\bibitem{pal2022blockchain}
S.~Pal, A.~Dorri, and R.~Jurdak, ``{Blockchain for IoT access control: Recent
  trends and future research directions},'' \emph{Journal of Network and
  Computer Applications}, p. 103371, 2022.

\bibitem{riabi2019survey}
I.~Riabi, H.~K.~B. Ayed, and L.~A. Saidane, ``{A survey on blockchain based
  access control for Internet of Things},'' in \emph{2019 15th International
  Wireless Communications \& Mobile Computing Conference}.\hskip 1em plus 0.5em
  minus 0.4em\relax IEEE, 2019, pp. 502--507.

\bibitem{khan2022authorization}
A.~Khan, A.~Ahmad, M.~Ahmed, J.~Sessa, and M.~Anisetti, ``Authorization schemes
  for {Internet of Things}: requirements, weaknesses, future challenges and
  trends,'' \emph{Complex \& Intelligent Systems}, vol.~8, no.~5, pp.
  3919--3941, 2022.

\bibitem{drame2021centralized}
S.~Dram{\'e}-Maign{\'e}, M.~Laurent, L.~Castillo, and H.~Ganem, ``Centralized,
  distributed, and everything in between: Reviewing access control solutions
  for the {IoT},'' \emph{ACM Computing Surveys}, vol.~54, no.~7, pp. 1--34,
  2021.

\bibitem{ravidas2019access}
S.~Ravidas, A.~Lekidis, F.~Paci, and N.~Zannone, ``{Access control in
  Internet-of-Things: A survey},'' \emph{Journal of Network and Computer
  Applications}, vol. 144, pp. 79--101, 2019.

\bibitem{khalid2021survey}
T.~Khalid, M.~A.~K. Abbasi, M.~Zuraiz, A.~N. Khan, M.~Ali, R.~W. Ahmad, J.~J.
  Rodrigues, and M.~Aslam, ``A survey on privacy and access control schemes in
  fog computing,'' \emph{International Journal of Communication Systems},
  vol.~34, no.~2, p. e4181, 2021.

\bibitem{liu2019survey}
D.~Liu, Z.~Yan, W.~Ding, and M.~Atiquzzaman, ``A survey on secure data
  analytics in edge computing,'' \emph{IEEE Internet of Things Journal},
  vol.~6, no.~3, pp. 4946--4967, 2019.

\bibitem{grunbacher2003posix}
A.~Gr{\"u}nbacher, ``{POSIX Access Control Lists on Linux},'' in \emph{USENIX
  Annual Technical Conference, FREENIX Track}, vol. 259272, 2003.

\bibitem{huang2020attribute}
D.~Huang, Q.~Dong, and Y.~Zhu, \emph{Attribute-based encryption and access
  control}.\hskip 1em plus 0.5em minus 0.4em\relax CRC Press, 2020.

\bibitem{otto2022designing}
B.~Otto, M.~ten Hompel, and S.~Wrobel, \emph{Designing data spaces: The
  ecosystem approach to competitive advantage}.\hskip 1em plus 0.5em minus
  0.4em\relax Springer Nature, 2022.

\bibitem{dabbagh2021survey}
M.~Dabbagh, K.-K.~R. Choo, A.~Beheshti, M.~Tahir, and N.~S. Safa, ``A survey of
  empirical performance evaluation of permissioned blockchain platforms:
  Challenges and opportunities,'' \emph{Computers \& Security}, vol. 100, p.
  102078, 2021.

\bibitem{zhang2021resource}
L.~Zhang, Y.~Zou, W.~Wang, Z.~Jin, Y.~Su, and H.~Chen, ``Resource allocation
  and trust computing for blockchain-enabled edge computing system,''
  \emph{Computers \& Security}, vol. 105, p. 102249, 2021.

\bibitem{zheng2018blockchain}
Z.~Zheng, S.~Xie, H.-N. Dai, X.~Chen, and H.~Wang, ``Blockchain challenges and
  opportunities: A survey,'' \emph{International Journal of Web and Grid
  Services}, vol.~14, no.~4, pp. 352--375, 2018.

\bibitem{nakamoto2008bitcoin}
S.~Nakamoto, ``Bitcoin: A peer-to-peer electronic cash system,''
  \emph{Decentralized Business Review}, 2008.

\bibitem{wood2014ethereum}
G.~Wood \emph{et~al.}, ``Ethereum: A secure decentralised generalised
  transaction ledger,'' \emph{Ethereum Project Yellow Paper}, vol. 151, no.
  2014, pp. 1--32, 2014.

\bibitem{androulaki2018hyperledger}
E.~Androulaki, A.~Barger, V.~Bortnikov, C.~Cachin, K.~Christidis, A.~De~Caro,
  D.~Enyeart, C.~Ferris, G.~Laventman, Y.~Manevich \emph{et~al.},
  ``{Hyperledger Fabric: A distributed operating system for permissioned
  blockchains},'' in \emph{Proceedings of the Thirteenth EuroSys Conference},
  2018, pp. 1--15.

\bibitem{botta2016integration}
A.~Botta, W.~De~Donato, V.~Persico, and A.~Pescap{\'e}, ``Integration of cloud
  computing and internet of things: a survey,'' \emph{Future Generation
  Computer Systems}, vol.~56, pp. 684--700, 2016.

\bibitem{lim2020federated}
W.~Y.~B. Lim, N.~C. Luong, D.~T. Hoang, Y.~Jiao, Y.-C. Liang, Q.~Yang,
  D.~Niyato, and C.~Miao, ``Federated learning in mobile edge networks: A
  comprehensive survey,'' \emph{IEEE Communications Surveys \& Tutorials},
  vol.~22, no.~3, pp. 2031--2063, 2020.

\bibitem{varghese2016challenges}
B.~Varghese, N.~Wang, S.~Barbhuiya, P.~Kilpatrick, and D.~S. Nikolopoulos,
  ``Challenges and opportunities in edge computing,'' in \emph{2016 IEEE
  International Conference on Smart Cloud}.\hskip 1em plus 0.5em minus
  0.4em\relax IEEE, 2016, pp. 20--26.

\bibitem{ren2019survey}
J.~Ren, D.~Zhang, S.~He, Y.~Zhang, and T.~Li, ``A survey on end-edge-cloud
  orchestrated network computing paradigms: Transparent computing, mobile edge
  computing, fog computing, and cloudlet,'' \emph{ACM Computing Surveys},
  vol.~52, no.~6, pp. 1--36, 2019.

\bibitem{cao2020overview}
K.~Cao, Y.~Liu, G.~Meng, and Q.~Sun, ``An overview on edge computing
  research,'' \emph{IEEE Access}, vol.~8, pp. 85\,714--85\,728, 2020.

\bibitem{xiao2019edge}
Y.~Xiao, Y.~Jia, C.~Liu, X.~Cheng, J.~Yu, and W.~Lv, ``Edge computing security:
  State of the art and challenges,'' \emph{Proceedings of the IEEE}, vol. 107,
  no.~8, pp. 1608--1631, 2019.

\bibitem{khan2019edge}
W.~Z. Khan, E.~Ahmed, S.~Hakak, I.~Yaqoob, and A.~Ahmed, ``Edge computing: A
  survey,'' \emph{Future Generation Computer Systems}, vol.~97, pp. 219--235,
  2019.

\bibitem{liu2019dynamic}
C.-F. Liu, M.~Bennis, M.~Debbah, and H.~V. Poor, ``Dynamic task offloading and
  resource allocation for ultra-reliable low-latency edge computing,''
  \emph{IEEE Transactions on Communications}, vol.~67, no.~6, pp. 4132--4150,
  2019.

\bibitem{kai2020collaborative}
C.~Kai, H.~Zhou, Y.~Yi, and W.~Huang, ``Collaborative cloud-edge-end task
  offloading in mobile-edge computing networks with limited communication
  capability,'' \emph{IEEE Transactions on Cognitive Communications and
  Networking}, vol.~7, no.~2, pp. 624--634, 2020.

\bibitem{DLM}
``{The Importance of Data Lifecycle Management (DLM) and Best Practices},''
  \url{https://www.computer.org/publications/tech-news/trends/the-importance-of-data-lifecycle-management},
  2022.

\bibitem{wang2020convergence}
X.~Wang, Y.~Han, V.~C. Leung, D.~Niyato, X.~Yan, and X.~Chen, ``Convergence of
  edge computing and deep learning: A comprehensive survey,'' \emph{IEEE
  Communications Surveys \& Tutorials}, vol.~22, no.~2, pp. 869--904, 2020.

\bibitem{fan2021serving}
Z.~Fan, W.~Yang, F.~Wu, J.~Cao, and W.~Shi, ``Serving at the edge: An edge
  computing service architecture based on icn,'' \emph{ACM Transactions on
  Internet Technology}, vol.~22, no.~1, pp. 1--27, 2021.

\bibitem{bozorgchenani2018centralized}
A.~Bozorgchenani, D.~Tarchi, and G.~E. Corazza, ``Centralized and distributed
  architectures for energy and delay efficient fog network-based edge computing
  services,'' \emph{IEEE Transactions on Green Communications and Networking},
  vol.~3, no.~1, pp. 250--263, 2018.

\bibitem{ning2020distributed}
Z.~Ning, P.~Dong, X.~Wang, S.~Wang, X.~Hu, S.~Guo, T.~Qiu, B.~Hu, and R.~Y.
  Kwok, ``Distributed and dynamic service placement in pervasive edge computing
  networks,'' \emph{IEEE Transactions on Parallel and Distributed Systems},
  vol.~32, no.~6, pp. 1277--1292, 2020.

\bibitem{wang2020edge}
T.~Wang, L.~Qiu, A.~K. Sangaiah, A.~Liu, M.~Z.~A. Bhuiyan, and Y.~Ma,
  ``{Edge-computing-based trustworthy data collection model in the Internet of
  Things},'' \emph{IEEE Internet of Things Journal}, vol.~7, no.~5, pp.
  4218--4227, 2020.

\bibitem{di2007data}
S.~D.~C. Di~Vimercati, S.~Foresti, S.~Jajodia, S.~Paraboschi, and P.~Samarati,
  ``A data outsourcing architecture combining cryptography and access
  control,'' in \emph{Proceedings of the 2007 ACM Workshop on Computer Security
  Architecture}, 2007, pp. 63--69.

\bibitem{BlockchainforAC}
V.~C. Hu and V.~C. Hu, \emph{Blockchain for access control systems}.\hskip 1em
  plus 0.5em minus 0.4em\relax US Department of Commerce, National Institute of
  Standards and Technology, 2022.

\bibitem{grammatikis2019securing}
P.~I.~R. Grammatikis, P.~G. Sarigiannidis, and I.~D. Moscholios, ``{Securing
  the Internet of Things: Challenges, threats and solutions},'' \emph{Internet
  of Things}, vol.~5, pp. 41--70, 2019.

\bibitem{dammak2020decentralized}
M.~Dammak, S.-M. Senouci, M.~A. Messous, M.~H. Elhdhili, and C.~Gransart,
  ``{Decentralized lightweight group key management for dynamic access control
  in IoT environments},'' \emph{IEEE Transactions on Network and Service
  Management}, vol.~17, no.~3, pp. 1742--1757, 2020.

\bibitem{jiang2022trust}
B.~Jiang, G.~Huang, T.~Wang, J.~Gui, and X.~Zhu, ``Trust based energy efficient
  data collection with unmanned aerial vehicle in edge network,''
  \emph{Transactions on Emerging Telecommunications Technologies}, vol.~33,
  no.~6, p. e3942, 2022.

\bibitem{atlam2021fuzzy}
H.~F. Atlam, R.~J. Walters, G.~B. Wills, and J.~Daniel, ``{Fuzzy logic with
  expert judgment to implement an adaptive risk-based access control model for
  IoT},'' \emph{Mobile Networks and Applications}, pp. 1--13, 2021.

\bibitem{dougherty2021apecs}
S.~Dougherty, R.~Tourani, G.~Panwar, R.~Vishwanathan, S.~Misra, and
  S.~Srikanteswara, ``{APECS: A distributed access control framework for
  pervasive edge computing services},'' in \emph{Proceedings of the 2021 ACM
  SIGSAC Conference on Computer and Communications Security}, 2021, pp.
  1405--1420.

\bibitem{AADEC}
X.~Zhou, D.~He, J.~Ning, M.~Luo, and X.~Huang, ``{AADEC: Anonymous and
  auditable distributed access control for edge computing services},''
  \emph{IEEE Transactions on Information Forensics and Security}, pp. 1--1,
  2022.

\bibitem{liu2022secure}
L.~Liu, C.~Huang, D.~Zhu, D.~Liu, J.~Ni, and X.~S. Shen, ``Secure and
  distributed access control for dynamic pervasive edge computing services,''
  in \emph{GLOBECOM 2022-2022 IEEE Global Communications Conference}.\hskip 1em
  plus 0.5em minus 0.4em\relax IEEE, 2022, pp. 5487--5492.

\bibitem{alrawais2017attribute}
A.~Alrawais, A.~Alhothaily, C.~Hu, X.~Xing, and X.~Cheng, ``An attribute-based
  encryption scheme to secure fog communications,'' \emph{IEEE Access}, vol.~5,
  pp. 9131--9138, 2017.

\bibitem{zhang2022achieving}
C.~Zhang, M.~Zhao, Y.~Xu, T.~Wu, Y.~Li, L.~Zhu, and H.~Wang, ``Achieving fuzzy
  matching data sharing for secure cloud-edge communication,'' \emph{China
  Communications}, vol.~19, no.~7, pp. 257--276, 2022.

\bibitem{goransson2016software}
P.~Goransson, C.~Black, and T.~Culver, \emph{Software defined networks: A
  comprehensive approach}.\hskip 1em plus 0.5em minus 0.4em\relax Morgan
  Kaufmann, 2016.

\bibitem{oberko2022survey}
P.~S.~K. Oberko, V.-H. K.~S. Obeng, and H.~Xiong, ``A survey on multi-authority
  and decentralized attribute-based encryption,'' \emph{Journal of Ambient
  Intelligence and Humanized Computing}, pp. 1--19, 2022.

\bibitem{sarma2022macfi}
R.~Sarma, C.~Kumar, and F.~A. Barbhuiya, ``{MACFI: A multi-authority access
  control scheme with efficient ciphertext and secret key size for fog-enhanced
  IoT},'' \emph{Journal of Systems Architecture}, vol. 123, p. 102347, 2022.

\bibitem{xu2020match}
S.~Xu, J.~Ning, Y.~Li, Y.~Zhang, G.~Xu, X.~Huang, and R.~Deng, ``Match in my
  way: Fine-grained bilateral access control for secure cloud-fog computing,''
  \emph{IEEE Transactions on Dependable and Secure Computing}, 2020.

\bibitem{cheng2021efficient}
R.~Cheng, K.~Wu, Y.~Su, W.~Li, W.~Cui, and J.~Tong, ``{An efficient ECC-based
  CP-ABE scheme for power IoT},'' \emph{Processes}, vol.~9, no.~7, p. 1176,
  2021.

\bibitem{xu2021server}
S.~Xu, J.~Ning, X.~Huang, J.~Zhou, and R.~H. Deng, ``Server-aided bilateral
  access control for secure data sharing with dynamic user groups,'' \emph{IEEE
  Transactions on Information Forensics and Security}, vol.~16, pp. 4746--4761,
  2021.

\bibitem{CAABAC}
A.~Arfaoui, S.~Cherkaoui, A.~Kribeche, and S.~M. Senouci, ``{Context-aware
  adaptive remote access for IoT applications},'' \emph{IEEE Internet of Things
  Journal}, vol.~7, no.~1, pp. 786--799, 2020.

\bibitem{ghosh2021case}
T.~Ghosh, A.~Roy, S.~Misra, and N.~S. Raghuwanshi, ``{CASE: A context-aware
  security scheme for preserving data privacy in IoT-enabled society 5.0},''
  \emph{IEEE Internet of Things Journal}, vol.~9, no.~4, pp. 2497--2504, 2021.

\bibitem{zheng2022adaptive}
W.~Zheng, B.~Chen, and D.~He, ``An adaptive access control scheme based on
  trust degrees for edge computing,'' \emph{Computer Standards \& Interfaces},
  vol.~82, p. 103640, 2022.

\bibitem{OpenVDAP}
Q.~Zhang, Y.~Wang, X.~Zhang, L.~Liu, X.~Wu, W.~Shi, and H.~Zhong, ``{OpenVDAP}:
  An open vehicular data analytics platform for {CAVs},'' in \emph{2018 IEEE
  38th International Conference on Distributed Computing Systems}, 2018, pp.
  1310--1320.

\bibitem{AC4AV}
Q.~Zhang, H.~Zhong, J.~Cui, L.~Ren, and W.~Shi, ``{AC4AV}: A flexible and
  dynamic access control framework for connected and autonomous vehicles,''
  \emph{IEEE Internet of Things Journal}, vol.~8, no.~3, pp. 1946--1958, 2021.

\bibitem{sun2022practical}
J.~Sun, G.~Xu, T.~Zhang, M.~Alazab, and R.~H. Deng, ``A practical fog-based
  privacy-preserving online car-hailing service system,'' \emph{IEEE
  Transactions on Information Forensics and Security}, vol.~17, pp. 2862--2877,
  2022.

\bibitem{sun2022secure}
J.~Sun, G.~Xu, T.~Zhang, X.~Cheng, X.~Han, and M.~Tang, ``Secure data sharing
  with flexible cross-domain authorization in autonomous vehicle systems,''
  \emph{IEEE Transactions on Intelligent Transportation Systems}, 2022.

\bibitem{9830119}
Y.~Bao, W.~Qiu, X.~Cheng, and J.~Sun, ``Fine-grained data sharing with enhanced
  privacy protection and dynamic users group service for the {IoV},''
  \emph{IEEE Transactions on Intelligent Transportation Systems}, pp. 1--15,
  2022.

\bibitem{zhang2021lightweight}
A.~Zhang, X.~Wang, X.~Ye, and X.~Xie, ``Lightweight and fine-grained access
  control for cloud--fog-based electronic medical record sharing systems,''
  \emph{International Journal of Communication Systems}, vol.~34, no.~13, p.
  e4909, 2021.

\bibitem{nasiraee2021privacy}
H.~Nasiraee and M.~Ashouri-Talouki, ``Privacy-preserving distributed data
  access control for {CloudIoT},'' \emph{IEEE Transactions on Dependable and
  Secure Computing}, 2021.

\bibitem{gaj2009fpga}
K.~Gaj and P.~Chodowiec, ``Fpga and asic implementations of aes,''
  \emph{Cryptographic engineering}, pp. 235--294, 2009.

\bibitem{EssentialsofEC}
``Essentials of edge computing.''
  \url{https://www.nxp.com/document/guide/essentials-of-edge-computing-ebook:EDGE-COMPUTING-EBOOK}.

\bibitem{cao2020policy}
Q.~H. Cao, M.~Giyyarpuram, R.~Farahbakhsh, and N.~Crespi, ``Policy-based usage
  control for a trustworthy data sharing platform in smart cities,''
  \emph{Future Generation Computer Systems}, vol. 107, pp. 998--1010, 2020.

\bibitem{munoz2020data}
A.~Munoz-Arcentales, S.~L{\'o}pez-Pernas, A.~Pozo, {\'A}.~Alonso,
  J.~Salvach{\'u}a, and G.~Huecas, ``Data usage and access control in
  industrial data spaces: Implementation using fiware,'' \emph{Sustainability},
  vol.~12, no.~9, p. 3885, 2020.

\bibitem{kelbert2018data}
F.~Kelbert and A.~Pretschner, ``Data usage control for distributed systems,''
  \emph{ACM Transactions on Privacy and Security}, vol.~21, no.~3, pp. 1--32,
  2018.

\bibitem{cirillo2020intentkeeper}
F.~Cirillo, B.~Cheng, R.~Porcellana, M.~Russo, G.~Solmaz, H.~Sakamoto, and
  S.~P. Romano, ``{IntentKeeper: Intent-oriented data usage control for
  federated data analytics},'' in \emph{2020 IEEE 45th Conference on Local
  Computer Networks}.\hskip 1em plus 0.5em minus 0.4em\relax IEEE, 2020, pp.
  204--215.

\bibitem{xue2022sparkac}
T.~Xue, Y.~Wen, B.~Luo, G.~Li, Y.~Li, B.~Zhang, Y.~Zheng, Y.~Hu, and D.~Meng,
  ``{SparkAC: Fine-grained access control in Spark for secure data sharing and
  analytics},'' \emph{IEEE Transactions on Dependable and Secure Computing},
  2022.

\bibitem{arora2022higher}
C.~Arora, S.~Z.~R. Rizvi, and P.~W. Fong, ``{Higher-order relationship-based
  access control: A temporal instantiation with IoT applications},'' in
  \emph{Proceedings of the 27th ACM on Symposium on Access Control Models and
  Technologies}, 2022, pp. 223--234.

\bibitem{cheng2017fogflow}
B.~Cheng, G.~Solmaz, F.~Cirillo, E.~Kovacs, K.~Terasawa, and A.~Kitazawa,
  ``{FogFlow: Easy programming of IoT services over cloud and edges for smart
  cities},'' \emph{IEEE Internet of Things Journal}, vol.~5, no.~2, pp.
  696--707, 2017.

\bibitem{ODRL}
``{ODRL Information Model 2.2},'' \url{https://www.w3.org/TR/odrl-model/}.

\bibitem{byun2008purpose}
J.-W. Byun and N.~Li, ``Purpose based access control for privacy protection in
  relational database systems,'' \emph{The VLDB Journal}, vol.~17, no.~4, pp.
  603--619, 2008.

\bibitem{liu2020fabric}
H.~Liu, D.~Han, and D.~Li, ``{Fabric-IoT: A blockchain-based access control
  system in IoT},'' \emph{IEEE Access}, vol.~8, pp. 18\,207--18\,218, 2020.

\bibitem{mazzocca2022framh}
C.~Mazzocca, N.~Romandini, M.~Colajanni, and R.~Montanari, ``{FRAMH: A
  federated learning risk-based authorization middleware for healthcare},''
  \emph{IEEE Transactions on Computational Social Systems}, 2022.

\bibitem{sylla2021blockchain}
T.~Sylla, L.~Mendiboure, M.~A. Chalouf, and F.~Krief, ``{Blockchain-based
  context-aware authorization management as a service in IoT},''
  \emph{Sensors}, vol.~21, no.~22, p. 7656, 2021.

\bibitem{liu2021capability}
Y.~Liu, Q.~Lu, S.~Chen, Q.~Qu, H.~O’Connor, K.-K.~R. Choo, and H.~Zhang,
  ``{Capability-based IoT access control using blockchain},'' \emph{Digital
  Communications and Networks}, vol.~7, no.~4, pp. 463--469, 2021.

\bibitem{xu2019exploration}
R.~Xu, Y.~Chen, E.~Blasch, and G.~Chen, ``Exploration of blockchain-enabled
  decentralized capability-based access control strategy for space situation
  awareness,'' \emph{Optical Engineering}, vol.~58, no.~4, pp.
  041\,609--041\,609, 2019.

\bibitem{xu2018blendcac}
------, ``{BlendCAC: A smart contract enabled decentralized capability-based
  access control mechanism for the IoT},'' \emph{Computers}, vol.~7, no.~3,
  p.~39, 2018.

\bibitem{bouras2021iot}
M.~A. Bouras, B.~Xia, A.~O. Abuassba, H.~Ning, and Q.~Lu, ``{IoT-CCAC: a
  blockchain-based consortium capability access control approach for IoT},''
  \emph{PeerJ Computer Science}, vol.~7, p. e455, 2021.

\bibitem{nakamura2020exploiting}
Y.~Nakamura, Y.~Zhang, M.~Sasabe, and S.~Kasahara, ``{Exploiting smart
  contracts for capability-based access control in the Internet of Things},''
  \emph{Sensors}, vol.~20, no.~6, p. 1793, 2020.

\bibitem{chen2023capability}
Y.~Chen, L.~Tao, B.~Liang, L.~Sun, Y.~Li, B.~Xing, and L.~Chen,
  ``{Capability-\& Blockchain-based fine-grained and flexible access control
  model},'' \emph{IEEE Network}, 2023.

\bibitem{zhang2021lightweight1}
J.~Zhang, L.~Yuan, and S.~Xu, ``A lightweight blockchain-based access control
  scheme for integrated edge computing in the {Internet of Things},''
  \emph{arXiv preprint arXiv:2111.06544}, 2021.

\bibitem{yu2020enabling}
G.~Yu, X.~Zha, X.~Wang, W.~Ni, K.~Yu, P.~Yu, J.~A. Zhang, R.~P. Liu, and Y.~J.
  Guo, ``Enabling attribute revocation for fine-grained access control in
  {blockchain-IoT} systems,'' \emph{IEEE Transactions on Engineering
  Management}, vol.~67, no.~4, pp. 1213--1230, 2020.

\bibitem{han2021blockchain}
D.~Han, Y.~Zhu, D.~Li, W.~Liang, A.~Souri, and K.-C. Li, ``A blockchain-based
  auditable access control system for private data in service-centric {IoT}
  environments,'' \emph{IEEE Transactions on Industrial Informatics}, vol.~18,
  no.~5, pp. 3530--3540, 2021.

\bibitem{yang2020secure}
W.~Yang, Z.~Guan, L.~Wu, X.~Du, and M.~Guizani, ``Secure data access control
  with fair accountability in smart grid data sharing: An edge blockchain
  approach,'' \emph{IEEE Internet of Things Journal}, vol.~8, no.~10, pp.
  8632--8643, 2020.

\bibitem{syed2020novel}
T.~A. Syed, M.~S. Siddique, A.~Nadeem, A.~Alzahrani, S.~Jan, and M.~A.~K.
  Khattak, ``A novel blockchain-based framework for vehicle life cycle
  tracking: An end-to-end solution,'' \emph{IEEE Access}, vol.~8, pp.
  111\,042--111\,063, 2020.

\bibitem{9662435}
G.~Wu, S.~Wang, Z.~Ning, and J.~L. records, ``Blockchain-enabled
  privacy-preserving access control for data publishing and sharing in the
  {Internet of Medical Things},'' \emph{IEEE Internet of Things Journal},
  vol.~9, no.~11, pp. 8091--8104, June 2022.

\bibitem{zhaofeng2019blockchain}
M.~Zhaofeng, W.~Lingyun, W.~Xiaochang, W.~Zhen, and Z.~Weizhe,
  ``Blockchain-enabled decentralized trust management and secure usage control
  of {IoT} big data,'' \emph{IEEE Internet of Things Journal}, vol.~7, no.~5,
  pp. 4000--4015, 2019.

\bibitem{xiao2020privacyguard}
Y.~Xiao, N.~Zhang, J.~Li, W.~Lou, and Y.~T. Hou, ``{PrivacyGuard: Enforcing
  private data usage control with blockchain and attested off-chain contract
  execution},'' in \emph{Computer Security--ESORICS 2020: 25th European
  Symposium on Research in Computer Security, ESORICS 2020, Guildford, UK,
  September 14--18, 2020, Proceedings, Part II 25}.\hskip 1em plus 0.5em minus
  0.4em\relax Springer, 2020, pp. 610--629.

\bibitem{zhang2021data}
X.~Zhang, X.~Li, Y.~Miao, X.~Luo, Y.~Wang, S.~Ma, and J.~Weng, ``A data trading
  scheme with efficient data usage control for industrial {IoT},'' \emph{IEEE
  Transactions on Industrial Informatics}, vol.~18, no.~7, pp. 4456--4465,
  2021.

\bibitem{pol2021preserving}
T.~Pol and A.~Badescu, ``{Preserving the privacy of data in autonomous cars IoT
  using Intel SGX},'' \emph{19th SC@ RUG 2021-2022}, p.~27.

\bibitem{gao2021blockchain}
Y.~Gao, H.~Lin, Y.~Chen, and Y.~Liu, ``{Blockchain and SGX-enabled
  edge-computing-empowered secure IoMT data analysis},'' \emph{IEEE Internet of
  Things Journal}, vol.~8, no.~21, pp. 15\,785--15\,795, 2021.

\bibitem{han2022blockchain}
J.~Han, Y.~Zhang, J.~Liu, Z.~Li, M.~Xian, H.~Wang, F.~Mao, and Y.~Chen, ``A
  blockchain-based and {SGX}-enabled access control framework for {IoT},''
  \emph{Electronics}, vol.~11, no.~17, p. 2710, 2022.

\bibitem{international2018new}
I.~B.~M. Corporation, ``Why new off-chain storage is required for
  blockchains,'' 2018.

\bibitem{zhu2022fine}
Y.~Zhu, X.~Wu, and Z.~Hu, ``Fine grained access control based on smart contract
  for edge computing,'' \emph{Electronics}, vol.~11, no.~1, p. 167, 2022.

\bibitem{ouaddah2016fairaccess}
A.~Ouaddah, A.~Abou~Elkalam, and A.~Ait~Ouahman, ``{FairAccess: a new
  blockchain-based access control framework for the Internet of Things},''
  \emph{Security and Communication Networks}, vol.~9, no.~18, pp. 5943--5964,
  2016.

\bibitem{ouaddah2021fairaccess2}
A.~Ouaddah and B.~Bellaj, ``{FairAccess2.0: a smart contract-based
  authorisation framework for enabling granular access control in IoT},''
  \emph{International Journal of Information and Computer Security}, vol.~15,
  no.~1, pp. 18--48, 2021.

\bibitem{novo2018blockchain}
O.~Novo, ``{Blockchain meets IoT: An architecture for scalable access
  management in IoT},'' \emph{IEEE Internet of Things Journal}, vol.~5, no.~2,
  pp. 1184--1195, 2018.

\bibitem{zhang2018smart}
Y.~Zhang \emph{et~al.}, ``{Smart contract-based access control for the Internet
  of Things},'' \emph{IEEE Internet of Things Journal}, vol.~6, no.~2, pp.
  1594--1605, 2018.

\bibitem{nobi2022towards}
M.~N. Nobi, ``{Towards machine learning based access control},'' Ph.D.
  dissertation, The University of Texas at San Antonio, 2022.

\bibitem{narouei2018automatic}
M.~Narouei, H.~Takabi, and R.~Nielsen, ``Automatic extraction of access control
  policies from natural language documents,'' \emph{IEEE Transactions on
  Dependable and Secure Computing}, vol.~17, no.~3, pp. 506--517, 2018.

\bibitem{ait2017abac}
M.~Ait El~Hadj, Y.~Benkaouz, B.~Freisleben, and M.~Erradi, ``{ABAC rule
  reduction via similarity computation},'' in \emph{Networked Systems: 5th
  International Conference, NETYS 2017, Marrakech, Morocco, May 17-19, 2017,
  Proceedings 5}.\hskip 1em plus 0.5em minus 0.4em\relax Springer, 2017, pp.
  86--100.

\bibitem{xiao2012automated}
X.~Xiao, A.~Paradkar, S.~Thummalapenta, and T.~Xie, ``Automated extraction of
  security policies from natural-language software documents,'' in
  \emph{Proceedings of the ACM SIGSOFT 20th International Symposium on the
  Foundations of Software Engineering}, 2012, pp. 1--11.

\bibitem{hosseinzadeh2020systematic}
A.~Hosseinzadeh, A.~Eitel, and C.~Jung, ``A systematic approach toward
  extracting technically enforceable policies from data usage control
  requirements.'' in \emph{International Conference on Information Systems
  Security and Privacy}, 2020, pp. 397--405.

\bibitem{nobi2022machine}
M.~N. Nobi, M.~Gupta, L.~Praharaj, M.~Abdelsalam, R.~Krishnan, and R.~Sandhu,
  ``Machine learning in access control: A taxonomy and survey,'' \emph{arXiv
  preprint arXiv:2207.01739}, 2022.

\bibitem{achiam2023gpt}
J.~Achiam, S.~Adler, S.~Agarwal, L.~Ahmad, I.~Akkaya, F.~L. Aleman, D.~Almeida,
  J.~Altenschmidt, S.~Altman, S.~Anadkat \emph{et~al.}, ``Gpt-4 technical
  report,'' \emph{arXiv preprint arXiv:2303.08774}, 2023.

\bibitem{abu2023flap}
A.~Abu~Jabal, E.~Bertino, J.~Lobo, D.~Verma, S.~Calo, and A.~Russo, ``Flap-a
  federated learning framework for attribute-based access control policies,''
  in \emph{Proceedings of the Thirteenth ACM Conference on Data and Application
  Security and Privacy}, 2023, pp. 263--272.

\bibitem{yu2020learning}
T.~Yu, T.~Li, Y.~Sun, S.~Nanda, V.~Smith, V.~Sekar, and S.~Seshan, ``Learning
  context-aware policies from multiple smart homes via federated multi-task
  learning,'' in \emph{2020 IEEE/ACM Fifth International Conference on
  Internet-of-Things Design and Implementation}.\hskip 1em plus 0.5em minus
  0.4em\relax IEEE, 2020, pp. 104--115.

\bibitem{nobi2022adversarial}
M.~N. Nobi, R.~Krishnan, and R.~Sandhu, ``Adversarial attacks in machine
  learning based access control,'' 2022.

\bibitem{pal2021protocol}
S.~Pal and Z.~Jadidi, ``{Protocol-based and hybrid access control for the IoT:
  approaches and research opportunities},'' \emph{Sensors}, vol.~21, no.~20, p.
  6832, 2021.

\bibitem{saha2021dhacs}
R.~Saha, G.~Kumar, M.~Conti, T.~Devgun, T.-h. Kim, M.~Alazab, and R.~Thomas,
  ``{DHACS: Smart contract-based decentralized hybrid access control for
  Industrial Internet-of-Things},'' \emph{IEEE Transactions on Industrial
  Informatics}, vol.~18, no.~5, pp. 3452--3461, 2021.

\bibitem{mahmud2022ifogsim2}
R.~Mahmud, S.~Pallewatta, M.~Goudarzi, and R.~Buyya, ``{iFogSim2}: An extended
  ifogsim simulator for mobility, clustering, and microservice management in
  edge and fog computing environments,'' \emph{Journal of Systems and
  Software}, vol. 190, p. 111351, 2022.

\bibitem{nandan2019iotsim}
D.~Nandan~Jha, K.~Alwasel, A.~Alshoshan, X.~Huang, R.~K. Naha, S.~K. Battula,
  S.~Garg, D.~Puthal, P.~James, A.~Y. Zomaya \emph{et~al.}, ``{IoTSim-Edge: A
  simulation framework for modeling the behavior of IoT and edge computing
  environments},'' \emph{arXiv e-prints}, pp. arXiv--1910, 2019.

\bibitem{sonmez2018edgecloudsim}
C.~Sonmez, A.~Ozgovde, and C.~Ersoy, ``{EdgeCloudSim: An environment for
  performance evaluation of edge computing systems},'' \emph{Transactions on
  Emerging Telecommunications Technologies}, vol.~29, no.~11, p. e3493, 2018.

\end{thebibliography}

\end{document}